\documentclass[12pt]{article}
\pdfoutput=1
\usepackage{setspace, graphicx, amssymb, amsmath, latexsym, amsfonts, amscd, amsthm, multirow, tikz, ctable, mathdots,caption,array,diagbox,mathtools,float,subfig}
\usepackage{tabularx,cite,mathrsfs}
\usepackage{authblk}
\usepackage{color}

\usepackage[margin=2cm]{caption}

\usepackage{fullpage}
\usepackage{stmaryrd}
\usepackage{rotating}

\usepackage{hyperref}

\def\bR {\mathbb{R}}
\def\bZ {\mathbb{Z}}

\def\bP{\mathbb{P}}

\newcommand{\be}{\begin{equation}}
\newcommand{\ee}{\end{equation}}
\newcommand{\bea}{\begin{align}}
\newcommand{\eea}{\end{align}}
\newcommand{\nn}{\nonumber}

\def\({\left(}
\def\){\right)}

\newcommand{\Tr}{{\rm Tr \,}}

\numberwithin{equation}{section}
\def\ie{\begin{equation}\begin{aligned}}
\def\fe{\end{aligned}\end{equation}}

\newcommand{\preprint}[1]{\begin{table}[t]  
             \begin{flushright}               
             \hfill{\tt #1}                           
             \end{flushright}                 
             \end{table}}                    
\renewcommand{\title}[1]{\vbox{\center\LARGE{#1}}\vspace{5mm}}
\renewcommand{\author}[1]{\vbox{\center#1}\vspace{5mm}}
\newcommand{\address}[1]{\vbox{\center\em#1}}
\newcommand{\email}[1]{\vbox{\center\tt#1}\vspace{5mm}}

\begin{document}

\begin{titlepage}

\preprint{CALT-TH-2020-015}

\begin{center}

\hfill \\
\hfill \\
\vskip 1cm

\title{
Lessons from the Ramond sector
}

\author{Nathan Benjamin$^{a}$, Ying-Hsuan Lin$^b$}

\address{${}^a$Princeton Center for Theoretical Science, 
\\ 
Princeton University, Princeton, NJ 08544, USA
\\
~
\\
${}^b$Walter Burke Institute for Theoretical Physics,
\\ 
California Institute of Technology, Pasadena, CA 91125, USA
\\
~
\\
}

\email{nathanb@princeton.edu, yhlin@caltech.edu}

\end{center}

\vfill

\abstract{

We revisit the consistency of torus partition functions in (1+1)$d$ fermionic conformal field theories, combining old ingredients of modular invariance/covariance with a modernized understanding of bosonization/fermionization dualities.
Various lessons can be learned by simply examining the oft-ignored Ramond sector.
For several extremal/kinky modular functions in the bootstrap literature, we can either rule out or identify the underlying theory. 
We also revisit the ${\cal N} = 1$ Maloney-Witten partition function by calculating the spectrum in the Ramond sector, and further extending it to include the modular sum of seed Ramond characters.
Finally, we perform the full ${\cal N} = 1$ RNS modular bootstrap and obtain new universal results on the existence of relevant deformations preserving different amounts of supersymmetry.

}

\vfill

\end{titlepage}

\tableofcontents

\section{Introduction}

The conformal bootstrap program in (1+1)$d$ proved extremely powerful in the study and classification of rational theories \cite{Belavin:1984vu, Friedan:1983xq}.  Although there has been little progress towards a classification of \emph{irrational} theories, modular invariance is still a very powerful constraint. For instance, a universal formula was obtained by Cardy \cite{Cardy:1986ie} showing that the high energy limit of (1+1)$d$ conformal field theory (CFT) is dictated by properties of the vacuum (and low lying states).  Hellerman then demonstrated in \cite{Hellerman:2009bu} that modular constraints could not only produce asymptotic statements but also universal bounds applicable at intermediate energies.  Subsequent explorations, fueled by new analytic ideas and powerful numerical methods, led to an explosion of exciting new results \cite{Keller:2012mr, Friedan:2013cba, Qualls:2013eha, Keller:2014xba, Benjamin:2016fhe, Kraus:2016nwo, Collier:2016cls, Collier:2017shs, Bae:2017kcl, Dyer:2017rul, Anous:2018hjh, Bae:2018qym, Afkhami-Jeddi:2019zci, Lin:2019kpn, Mukhametzhanov:2019pzy, Hartman:2019pcd, Pal:2019zzr, Pal:2020aa}.

One qualitative difference between the modular bootstrap and the four-point correlator bootstrap is that the degeneracies of all states must be non-negative integers. It is generally not known how to efficiently implement this constraint, either numerically or analytically. However, given a putative CFT spectrum (for instance found at the extremal point of some semidefinite optimization procedure), it is easy to check if the coefficients are integers. In any case, given \emph{a} solution to the modular crossing equation with non-negative integer degeneracy, it is not guaranteed that such a spectrum can be realized by a compact, unitary CFT.

In this paper, we study consistency conditions for (1+1)$d$
fermionic CFTs.  In contrast to bosonic CFTs, fermionic CFTs must be defined on manifolds with a choice of the spin structure.\footnote{A prototypical example of a fermionic CFT is the (1+1)$d$ free Majorana fermion, which when defined on a higher genus Riemann surface requires specifying the periodicities about all cycles.  By contrast, its bosonization, the Ising CFT, can be defined even on non-orientable Riemann surfaces.
}
We consider two necessary (but not sufficient) consistency conditions on the spectra. The first concerns the relations among different sectors. For any fermionic theory, we can define two different Hilbert spaces with different 
periodicities for the fermions about the spatial circle: the Neveu-Schwarz (NS) sector and the Ramond (R) sector. Both sectors 
can be decomposed into modules 
expanded into characters 
of the algebra with non-negative integer coefficients. It could be the case that a perfectly healthy-looking partition function in the NS sector bears hidden sickness in the Ramond sector. For instance, in \cite{Bae:2018qym} the authors derived various universal bounds on $\mathcal{N}=(1,1)$ (and unextended $\mathcal{N}=(2,2)$) CFTs by studying the NS sector partition function; we extend this analysis to include the Ramond sector and find significantly stronger bounds. 

The second concerns the consistency condition on the chiral algebra of the CFT. Recall that for any fermionic CFT we can obtain a bosonic CFT via the GSO projection \cite{Gliozzi:1976qd}. At the level of the torus partition functions, this corresponds to summing over all four spin structures: periodic/anti-periodic boundary conditions on the space and time circles. This story has recently been modernized and enriched into more precise bosonization/fermionization maps \cite{Gaiotto:2015zta,Bhardwaj:2016clt,Kapustin:2017jrc,Thorngren:2018bhj,Gaiotto:2018ypj,Karch:2019lnn,yujitasi,Kaidi:2020aa,Ji:2019ugf,Hsin:2019gvb,Hsieh:2020aa,Kulp:2020aa}. Fermionic CFTs come in pairs related by tensoring with the fermionic symmetry-protected topological (SPT) phase $(-1)^{\text{Arf}}$ where Arf is the Arf invariant.\footnote{On a closed oriented Riemann surface with a given spin structure, the Arf invariant counts the number of zero modes of the Dirac operator mod 2.  For instance, on the torus, the Arf is 1 if the fermions are periodic around both cycles, and 0 otherwise.
}
In terms of the torus partition functions, the two theories share the same ones with even spin structures ($Z^{\text{NS}+}$, $Z^{\text{NS}-}$, $Z^{\text{R}+}$), and only differ by an overall sign on the one with odd spin structure ($Z^{\text{R}-}$). The corresponding bosonized CFTs are related by an orbifold with respect to the $\bZ_2$ global symmetry that is dual to fermion parity $(-1)^F$.
By analyzing consistency conditions on {\it both} bosonic CFTs obtained this way, we are able to constrain the possible spectra of the fermionic CFT.

As a proof of principle that both of these tests have teeth, we apply them to previous results in the literature. Using the first constraint, we are able to show that certain ``kinks" that had previously appeared in the numerical modular bootstrap analyses of \cite{Bae:2018qym} cannot correspond to physical compact, unitary CFTs; although the NS sector partition function naively looks healthy, the Ramond sector does not have integer degeneracies. Using the second constraint, we are able to explicitly identify the CFT associated with another kink that was found (but not identified) in \cite{Dyer:2017rul}.

This paper is organized as follows. In the remainder of the introduction, we will review salient facts about fermionic CFTs, including the different spin structures on torii and the corresponding partition functions. In Section \ref{Sec:Diagnostic} we will introduce our first diagnostic: positivity and integrality of degeneracies and unitarity of the Ramond sector. We will use this diagnostic to explore some candidate spectra, including previously-considered kinks from the modular bootstrap, and perform a more thorough analysis of the spectrum of a purported theory of pure supergravity. In Section \ref{sec:bosonizationbohnanza} we will first describe a Sugawara constraint on the chiral algebra of general bosonic CFTs, and then apply it to identify a specific fermionic CFT (via GSO projection). In Section~\ref{sec:lessons}, we use the numerical bootstrap to study general constraints on renormalization group flows. Finally we conclude in Section \ref{sec:conclude}.

\subsection{Fermionic partition functions and invariance subgroups} 

In a (1+1)$d$ fermionic CFT, there are four torus partition functions with different spin structures, which can be defined as traces over Hilbert spaces as 
\ie
\label{PartitionFunctions}
\renewcommand{\arraystretch}{1.5}
\begin{tabular}{m{8cm} m{2cm}}
$Z^{\text{R}-}(\tau, \bar\tau)
= \text{Tr}_{\cal H_\text{R}} [\,(-1)^Fq^{L_0-{c\over24}} \bar q^{\bar L_0-{c\over24}}\,]$ & $SL(2, \bZ)$
\\
$Z^{\text{R}+}(\tau, \bar\tau)
= \text{Tr}_{\cal H_\text{R}} [\, q^{L_0-{c\over24}} \bar q^{\bar L_0-{c\over24}}\,]$ & $\Gamma^0(2)$
\\
$Z^{\text{NS}-}(\tau, \bar\tau)
= \text{Tr}_{\cal H_\text{NS}} [\,\,(-1)^F q^{L_0-{c\over24}} \bar q^{\bar L_0-{c\over24}}\,]$ & $\Gamma_0(2)$
\\
$Z^{\text{NS}+}(\tau, \bar\tau)
= \text{Tr}_{\cal H_\text{NS}} [\,q^{L_0-{c\over24}} \bar q^{\bar L_0-{c\over24}}\,]$ & $\Gamma_\theta$
\end{tabular}
\fe
On the right, we specify the invariance subgroup of each partition function inside $SL(2, \bZ)$.
If we write the elements of $SL(2, \bZ)$ as
\ie
g = 
\begin{pmatrix}
a & b
\\
c & d
\end{pmatrix} \, , \quad a d - b c = 1 \, ,
\fe
then the invariance subgroups above are congruence subgroups defined as
\ie
& \Gamma^0(2) = \{ g \in SL(2, \bZ) ~|~ b \equiv 0 \mod 2 \} \, ,
\\
& \Gamma_0(2) = \{ g \in SL(2, \bZ) ~|~ c \equiv 0 \mod 2 \} \, ,
\\
& \Gamma_\theta ~\quad = \{ g \in SL(2, \bZ) ~|~a + d \equiv b+c \equiv 0 \mod 2 \} \, .
\fe
The latter three in \eqref{PartitionFunctions} transform as a modular vector
\ie
\begin{pmatrix}
Z^{\text{R}+}(\tau, \bar\tau)
\\
Z^\text{NS$-$}(\tau, \bar\tau)
\\
Z^\text{NS$+$}(\tau, \bar\tau)
\end{pmatrix} \, ,
\fe
with $S$ and $T$ matrices
\ie
S = 
\begin{pmatrix}
& 1
\\
1
\\
& & 1
\end{pmatrix} \, ,
\quad
T = 
\begin{pmatrix}
1
\\
& & 1
\\
& 1
\end{pmatrix} \, .
\fe
Given one of the three, the other two can be obtained by modular transformations.
By contrast, $Z^{\text R-}(\tau, \bar\tau)$ is invariant under all of $SL(2,\mathbb Z)$ and in general \emph{cannot} be determined from knowledge of any of the other three partition functions.  Nonetheless, from the trace definitions in \eqref{PartitionFunctions}, we know that $Z^{\text R-}(\tau, \bar\tau)$ consists of the same $q, \, \bar q$ powers as $Z^{\text R+}(\tau, \bar\tau)$, and with coefficients bounded by those of the latter.  Moreover, if the theory has $\mathcal{N}=(1,1)$ supersymmetry, then $Z^{\text{R}-}(\tau,\bar\tau)$ is a number, the Witten index.

\section{Diagnostics of $\Gamma_{\theta}$ modular functions}

Given a $\Gamma_{\theta}$ modular function admitting a decomposition into ${\cal N} = 1$ characters with positive integer degeneracies, can it be the NS+ partition function of a physical CFT?  
While we cannot answer this question in full generality, in this section we present two simple diagnostics that easily rule out some inconsistent $\Gamma_{\theta}$ modular functions.

The first is to examine the weights and degeneracies in the Ramond sector.  We find examples of extremal modular functions in the NS modular bootstrap literature that are unphysical in the Ramond sector under this criterion.  We also discuss the practicality of this diagnostic when the NS$+$ partition function is only known approximately.  

The second is to perform a GSO projection, and examine the consistency of the bosonized partition function.  
An immediate difficulty is that the R$-$ partition function cannot be determined from the knowledge of the NS+ partition function alone.  We first take a detour to point out a simple bound on the number of conserved currents by the Sugawara construction, and then impose it on the bosonized partition functions.  For a special extremal modular function found by bootstrap methods in \cite{Dyer:2017rul}, this consideration in fact guides us to actually identify the underlying CFT.

\subsection{Ramond sector unitarity and ground state degeneracy
}
\label{Sec:Diagnostic}

Given an NS+ partition function,
the Ramond sector partition function is given by
\be
\label{ST}
Z^{\text{R}+}(\tau, \bar\tau) = Z^{\text{NS}+}\(-\frac1\tau+1, -\frac1{\bar\tau}+1\) \, .
\ee
It may be the case that a $\Gamma_\theta$ modular function has a decomposition into NS characters with integer degeneracies and weights satisfying unitary bounds, but after \eqref{ST} has a problematic decomposition into Ramond characters.  

A simple diagnostic of the consistency of the Ramond sector is to check whether the ground state degeneracy is a non-negative integer, and whether the weights are above the unitarity bound.  This was used in Section~3.2 of \cite{Witten:2007kt} to rule out certain holomorphic $\mathcal{N}=1$ superconformal field theories\footnote{Patient 0 :p}  (and was further explored in Section 3.2 of \cite{Benjamin_2016}). In precise terms, the diagnostic requires
\be
\label{nRgs}
\lim_{\tau \to i\infty} q^{{c\over24}-h^\text{R}_\text{gs}} \bar q^{{\bar c\over24}-\bar h^\text{R}_\text{gs}} \, Z^{\text{R}+}(\tau, \bar\tau) = \lim_{\tau \to i\infty} q^{{c\over24}-h^\text{R}_\text{gs}} \bar q^{{\bar c\over24}-\bar h^\text{R}_\text{gs}} \, Z^{\text{NS}+}\(-\frac1\tau+1, -\frac1{\bar\tau}+1\) = n^\text{R}_\text{gs} \, ,
\ee
with $h^\text{R}_\text{gs}, \bar h^\text{R}_\text{gs} \ge 0$, to be a non-negative integer.
If we have ${\cal N} = (1,1)$ supersymmetry, then the unitarity bound further constrains $h^\text{R}_\text{gs}, \bar h^\text{R}_\text{gs} \ge {c\over24}$.  In particular, if
\be
\lim_{\tau \to i\infty} Z^{\text{R}+}(\tau, \bar\tau) = \lim_{\tau \to i\infty} Z^{\text{NS}+}\(-\frac1\tau+1, -\frac1{\bar\tau}+1\)
\ee
diverges, then it is in obvious conflict with ${\cal N} = (1,1)$ supersymmetry and unitarity.

It may be the case the modular function that requires diagnosis may only be known numerically or as a $q, \bar{q}$ series without an obvious closed form expression in terms of elementary modular functions.  This presents a difficulty in determining the Ramond sector partition function, since a modular $S$ transform is involved.  However, if we focus on the Ramond ground states as in \eqref{nRgs}, then the examination can even be done numerically as long as we can  approximate $Z^{\text{NS}+}(\tau, \bar\tau)$ well enough.  In practice, this may require knowledge of $Z^{\text{NS}+}(\tau, \bar\tau)$ to very high order in the $q, \bar{q}$-expansion.

In \cite{Bae:2018qym}, a modular bootstrap system of equations was analyzed for $\mathcal{N}=1$ SCFTs. However, they only analyzed the NS+ partition function and demanded that the partition function is well-behaved under $\Gamma_{\theta}$. 
We presently diagnose the two partition functions they found at $c=9$ and at $c=\frac{33}2$, which appeared as kinks in the numerical bounds, and will show that they cannot be the partition functions of physical CFTs.  
We then turn to the ${\cal N} = 1$ Maloney-Witten partition function \cite{Maloney:2007ud}. While the authors of \cite{Maloney:2007ud} computed the modular sum of the vacuum character in the NS sector, we extend their computation and analysis to the Ramond sector.

\subsubsection{Patient 1: Extremal ${\cal N} = 1$ NS$+$ modular function at $c=9$}

This 
partition function is given in (3.32) of \cite{Bae:2018qym}
\be
\label{c=9}
Z_{c=9}^{\text{NS}+}(\tau,\bar \tau) = \left |f_{c=9}^{\text{NS}+}(\tau)\right|^2
\ee
with
\ie
f_{c=9}^{\text{NS}+}(\tau) &= \(\frac{\eta(q)^{48}}{\eta(q^2)^{24}\eta(\sqrt{q})^{24}} - 18\)\(\frac{\eta(q^2)^6\eta(\sqrt q)^6}{\eta(q)^{12}}\)\\
&= q^{-\frac38} + 261 q^{\frac58} + 456 q^{\frac98} + 4500 q^{\frac{13}8} + \cdots \, .
\fe
By \eqref{ST}, and using the transformation properties of the Dedekind eta function, we find that
\begin{align}
Z_{c=9}^{\text{R}}(\tau, \bar\tau)= \left | f_{c=9}^\text R(\tau)\right |^2
\end{align}
with
\ie
f_{c=9}^{\text R}(\tau) &= \(\frac{2^{12}\eta(q^2)^{24}}{\eta(q)^{24}} + 18\)\(\frac{\eta(q)^6}{8\eta(q^2)^6}\) \\
&= \frac{9}{4q^{\frac14}} + \frac{997q^{\frac34}}{2} + \frac{36999q^{\frac74}}{4} + \cdots \, .
\label{eq:fc9R}
\fe
From (\ref{eq:fc9R}) we immediately see that the Ramond ground states have weights $(h, \bar h) =  \(\frac18, \frac18\)$, which violate the ${\cal N} = 1$ unitarity bound ($h, \bar h \ge \frac{c}{24} = \frac38$). More importantly, the degeneracies are not integers, so not even a non-supersymmetric fermionic CFT can have \eqref{c=9} as its NS+ partition function.\footnote{The sicknesses in Patients 1 and 2 were already pointed out in German in Table 5.4 of \cite{Hoehn}.\label{foot}}

\subsubsection{Patient 2: Extremal ${\cal N} = 1$ NS$+$ modular function at $c=\frac{33}2$}

This
partition function is given in (3.35) of \cite{Bae:2018qym}
\be
Z^{\text{NS}+}_{c={33\over2}}(\tau,\bar\tau) = \left| f_{c={33\over2}}^{\text{NS}+}(q) \right|^2
\ee
with
\be
f_{c={33\over2}}^{\text{NS}+}(q) = q^{-33/48}\(1 + 7766 q^{\frac32} + 11220 q^2 + \cdots\) \, .
\label{eq:fc332}
\ee 
The derivation of this partition function in \cite{Bae:2018qym} used the technique of modular differential equations, explained in their Appendix A.\footnote{The authors of \cite{Bae:2018qym} informed us that \eqref{eq:fc332} has an analytic expression in terms of the $c=\frac12$ free fermion partition function, using which one could analytically perform the modular $S$ transform and in particular verify \eqref{33gs}.
}
For completeness, we reproduce their analysis in Appendix \ref{sec:jaewon332}.
We computed \eqref{eq:fc332} to $q^{1000}$ and numerically found that at $\tau = 1+i\epsilon$ for small $\epsilon$,
\be
\label{33gs}
|f_{c={33\over2}}^{\text{R}}(q)| \sim \frac{231}{128\sqrt 2} e^{\frac{2\pi}{\epsilon}\(\frac{33}{48} - \frac1{16}\)} \, .
\ee
Thus, the putative CFT has Ramond ground state energy $\(\frac1{16}, \frac1{16}\)$ and degeneracy $\frac{53361}{32768}$. Again, this both violates the ${\cal N} = (1,1)$ unitarity bound and exhibits non-integer degeneracy.\footnote{See footnote \ref{foot}.} We conclude that no fermionic CFT can have this NS+ partition function.

\subsubsection{Patient 3: The $\mathcal N=1$ Maloney-Witten partition function}
\label{sec:mwkn1}

The (bosonic) Maloney-Witten partition function is a properly regularized sum of $SL(2,\mathbb Z)$ images of the Virasoro vacuum character \cite{Maloney:2007ud}. It is modular invariant by construction, and the spectrum has a gap at the BTZ black hole threshold, a feature sought in a putative theory of ``pure" gravity in AdS$_3$. Nevertheless, this partition function suffers the undesirable features of a continuous spectrum and a negative density of states for odd spins at low twists \cite{Benjamin:2019stq}.\footnote{Inspired by the relation between Jackiw-Teitelboim gravity and random matrix theory \cite{Saad:2019lba}, it has been speculated that the continuous spectrum of the Maloney-Witten partition function may mean that the gravity theory is dual to an ensemble averaged conformal field theory.
}

To produce an ${\cal N} = 1$ supersymmetric version of the Maloney-Witten partition function, one simply replaces the Virasoro vacuum character by the ${\cal N} = 1$ super-Virasoro vacuum character.\footnote{More generally, one could consider the vacuum character of any vertex operator algebra.
}
The original work of \cite{Maloney:2007ud} considered the NS sector, and computed\footnote{The sum over elements of $\Gamma^0(2)$ diverges, but can be appropriately regularized to give a finite sum. See \cite{Maloney:2007ud} for more details.}
\ie
Z^{\text{NS}-}(\tau,\bar\tau) &= \Tr_{\text{NS-NS}}\((-1)^F q^{L_0-\frac{c}{24}}\bar{q}^{\overline{L}_0-\frac c{24}}\) \\
&= \sum_{\gamma\in\Gamma^0(2)} \chi_0^{\text{NS}-}(\gamma\tau)\chi_0^{\text{NS}-}(\gamma\bar\tau) \, ,
\label{eq:Gamma02sum}
\fe
where $(-1)^F$ is $+1$ if $h-\bar{h} \in \mathbb Z$ and $-1$ if $h-\bar{h} \in \mathbb Z + \frac12$. The ${\cal N} = 1$ vacuum character $\chi_0^{\text{NS}-}(\tau)$ is given by
\be
\chi_0^{\text{NS}-}(\tau) = \frac{\eta(\tau/2)}{\eta(\tau)^2} q^{-\frac{c-\frac32}{24}}(1+q^{\frac12}) \, .
\ee
They noted the following:
\begin{enumerate}
\item The density of primaries in the NS sector is continuous for $h, \, \bar h > \frac{c-\frac32}{24}$.
\item The degeneracy of NS primaries at $h=\bar{h}=\frac{c-\frac32}{24}$ is negative and equal to $-6$. In \cite{Keller:2014xba}, it was pointed out that such negativities can be cured by adding free theories.  Here we add 6 times the partition function of the $c = {3\over2}$ free theory reviewed in Appendix~\ref{sec:c32ft}.
\end{enumerate}
In \cite{Benjamin:2019stq}, it was pointed out that the density of primary operators in the NS sector has an additional negativity:
\begin{enumerate}
\setcounter{enumi}{2}
\item The density of primaries in the NS sector is negative if the spin $h-\bar{h}$ is odd or half-integer, and the twist $2\,\text{min}(h, \bar h)$ is sufficiently close to $\frac{c-\frac32}{12}$. A cure is to add the modular sum of sufficiently many primary operators with twist $2\,\text{min}(h, \bar h) = \frac{c-\frac32}{16}$ (similar to the bosonic case).
\end{enumerate}
However, other aspects of the ${\cal N} = 1$ Maloney-Witten partition function have not been carefully examined, including the consistency of the Ramond sector spectrum. In addition, one could consider the modular sum of a single Ramond (as opposed to NS) character.
Appendix~\ref{sec:appmwk} is devoted to a detailed analysis of these aspects. The key discoveries are:
\begin{enumerate}
\setcounter{enumi}{3}
\item In the original Maloney-Witten sum of the NS vacuum character, the spectrum of the Ramond sector (obtained by a modular $S$ transform from the NS$-$) is continuous, lies above the unitarity bound $h, \, \bar h \ge {c\over24}$, and the ground state degeneracy is 16. The density of primaries in the Ramond sector is positive everywhere.
\item The modular sum of a non-degenerate NS character for a scalar primary produces $+2$ Ramond sector ground state degeneracy. 
\item The modular sum of the Ramond vacuum character {\it vanishes}.
\item The modular sum of a Ramond character for a scalar primary produces $-2$ Ramond ground state degeneracy.
\item All partition functions above can be written as sums of non-holomorphic Eisenstein series (see Appendix \ref{sec:appmwk}).
\end{enumerate}
Table~\ref{tab:rsector} summarizes the different contributions to ground state degeneracies.

\begin{table}[H]
	\begin{center}
		\renewcommand{\arraystretch}{1.5}
		\begin{tabular}{| c | c | c |}
			\hline
			& {\bf NS primary} & {\bf R ground state} \\
			{\bf Partition function} & {\bf degeneracy at} & {\bf degeneracy at}
			\\
			& {\bf $h=\bar{h} = \frac{c-\frac32}{24}$} & {\bf $h=\bar{h} = \frac{c}{24}$} \\
			\hline
			$\Gamma^0(2)$ sum of NS vacuum & $-6$ & $+16$ \\ \hline
			$c=\frac32$ free theory & $+1$ & $+2$ \\
			$\Gamma^0(2)$ sum of low-twist scalar NS primary & $-1$ & $0$ \\
			$\Gamma_0(2)$ sum of Ramond sector ground state & $0$ & $0$ \\
			$\Gamma_0(2)$ sum of Ramond sector scalar primary & $0$ & $-2$ \\ 
			\hline
		\end{tabular}
	\end{center}
	\caption{Contributions to degeneracies by different partition functions. The first row is the original Maloney-Witten partition function.  The rest can be added to cure the negativity in the density of primaries or to adjust the Ramond ground state degeneracy. 
	 }
	\label{tab:rsector}
\end{table}

A minimal combination of partition functions that does not appear to suffer any negativity is the sum of (i) the $\Gamma^0(2)$ modular sum of the $\mathcal{N}=1$ vacuum character, (ii) two copies of the $\Gamma^0(2)$ modular sum of an $\mathcal{N}=1$ scalar primary at $h=\bar h = \frac{c-\frac32}{32}$, and (iii) eight copies of the free $c=\frac32$ theory. This resulting partition function has a Ramond ground state degeneracy of $32$.

We note that every term in Table \ref{tab:rsector} has an even Ramond ground state degeneracy. If we insist on adding an integer-number multiple of every term in Table \ref{tab:rsector}, then so far we do not have a way to make the Ramond ground state degeneracy odd (or equivalently the Witten index odd). Although there exist supersymmetric theories with odd Witten index (for example the $\mathcal{N}=2$ minimal model $A$-series at $c=\frac{3k}{k+2}$ has Witten index $k+1$), we have not yet found a supersymmetric theory with odd Witten index that can be decomposed into non-degenerate $\mathcal{N}=1$ super-Virasoro characters with non-negative coefficients.\footnote{
The $c=\frac{3}{2}$ $\mathcal{N}=2$ minimal model is related to the $c=\frac{3}{2}$ free theory at $r=2$ (described in Appendix \ref{sec:c32ft}) in the following way: They are fermionizations of the same bosonic CFT --- the product of the Ising CFT with the $c=1$ free boson at $r = 2$ --- with respect to different $\bZ_2$ global symmetries.}

\subsection{Bosonization and the Sugawara construction}
\label{sec:bosonizationbohnanza}

\begin{figure}[H]
\centering
\subfloat{
\begin{tikzpicture}
\coordinate (Z2)  at (0,0);
\draw (4,0) node {$\cal B$};
\draw (4,-3) node {${\cal B}/\bZ_2$};
\draw (9,0) node {$\cal F$};
\draw (9.5,-3) node {${\cal F} \otimes (-1)^{\text{Arf}}$};
\draw [<-,thick] (5.5,0.1)--(7.4,0.1) node[midway,above] {bosonize};
\draw [->,thick] (5.5,- 0.1)--(7.4,-0.1) node[midway,below] {fermionize};
\draw  [<->, thick]  (4,-.6) -- node[left] {gauge $\mathbb{Z}_2$} (4,-2.5) ;
\draw  [<->, thick]  (9,-.6) -- node[right] {$\otimes~(-1)^{\text{Arf}}$} (9,-2.5);
\draw [<-,thick] (5.5,0.1-3) -- (7.4,0.1-3) node[midway,above] {bosonize};
\draw [->,thick] (5.5,- 0.1 - 3)--(7.4,-0.1-3) node[midway,below] {fermionize};
\end{tikzpicture}
}
\qquad
\subfloat{
\raisebox{.8in}{
\renewcommand{\arraystretch}{1.75}
\begin{tabular}{|c|c|}
\hline ${\cal B}$ & ~~~${\cal F}$~~~ \\\hline
${\cal H}_\text{even}$ & ${\cal H}^\text{NS}_\text{even}$ \\
${\cal H}_\text{odd}$ & ${\cal H}^\text{R}_\text{even}$ 
\\
${\cal H}^\text{defect}_\text{even}$ & ${\cal H}^\text{R}_\text{odd}$ 
\\
${\cal H}^\text{defect}_\text{odd}$ & ${\cal H}^\text{NS}_\text{odd}$
\\\hline 
\end{tabular}
}}
\quad~
\caption{The bosonization/fermionization map and the isomorphism of Hilbert spaces.  Even or odd refers to the charge under $(-1)^F$ or the dual $\bZ_2$.  On the left, defect means the defect Hilbert space quantized with twisted periodic boundary conditions (by $\bZ_2$). 
}\label{Fig:Square} 
\end{figure}
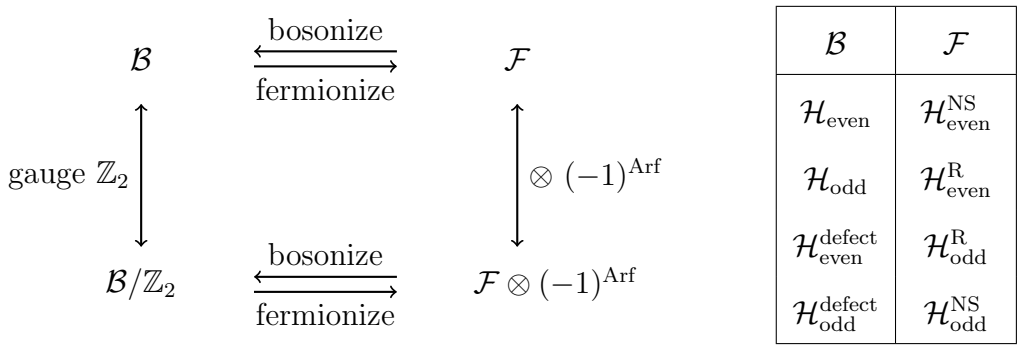

Bosonization
maps a fermionic CFT $\cal F$ to a bosonic CFT $\cal B$ with (dual) $\bZ_2$ global symmetry, {\it i.e.} GSO projection \cite{Gliozzi:1976qd}, and fermionization does the reverse.  On the one hand, $\cal B$ can be orbifolded to give rise to a different bosonic CFT ${\cal B}/\bZ_2$, and on the other hand, $\cal F$ can be tensored with the fermionic symmetry-protected topological (SPT) phase $(-1)^\text{Arf}$ to give rise to a different fermionic CFT ${\cal F} \otimes (-1)^\text{Arf}$. At the level of the torus partition functions, the latter maintains those with even spin structures ($Z^{\text{NS}+}$, $Z^{\text{NS}-}$, $Z^{\text{R}+}$), and flips the sign of the one with odd spin structure ($Z^{\text{R}-}$).\footnote{The same statement is true for partition functions on general Riemann surfaces.}  The result is a square of relations for a quadruple of CFTs shown in Figure~\ref{Fig:Square}.  We refer the reader to \cite{Karch:2019lnn,yujitasi,Kaidi:2020aa,Ji:2019ugf} for a pedagogical introduction to this subject.

Given a $\Gamma_\theta$ modular function, one could ask if its interpretation as the NS+ partition function of a fermionic CFT can be consistent with the square of maps shown in Figure~\ref{Fig:Square}.  We will impose this consistency condition to identity a previously unknown theory corresponding to an extremal $U(1)$ flavored modular function found in \cite{Dyer:2017rul}.
However, to do so, we must first take a detour: Section~\ref{Sec:Sugawara} presents a diagnostic for general bosonic partition functions, and is a self-contained argument independent of other parts of this paper. This diagnostic will then be used in Section~\ref{Sec:PatientDyer} in conjunction with bosonization/fermionization to study the extremal modular function of \cite{Dyer:2017rul}.

\subsubsection{Bounds on conserved currents from the Sugawara construction}

\label{Sec:Sugawara}

Ancient wisdom says that the stress tensor of any CFT can be decomposed into two parts
\ie
T = T_s + T_o \, ,
\fe
where $T_s$ is the Sugawara stress tensor of all spin-one conserved currents, and importantly the $T_s \, T_o$ OPE is {\it regular}.  By unitarity, the central charge $c$ of the CFT must be larger than the Sugawara central charge.
Furthermore, the number of spin-$s$ conserved currents for any $s \ge 2$ must be no smaller than the that furnished by the normal ordered products of spin-one conserved currents, since their norms are completely fixed by the singular part of the OPE of spin-one conserved currents.

We assume that the affine current algebra must be the tensor product of factors of $U(1)$ and a \emph{semi-simple} compact affine Lie algebras.\footnote{This excludes affine algebras whose global part contains a non-compact Lie algebra such as $SL(2, \bR)$ or an intermediate Lie algebra such as $E_{7{1\over2}}$\cite{Gel_fand_1985,Shtepin_1994,Landsberg:aa,Kawasetsu:2013}.
}
For a fixed number of spin-one conserved currents, there is a finite number of possibilities for the affine current algebra.  Enumerating over all the possibilities, and computing the corresponding Sugawara central charge, we derive universal lower bounds on the central charge and on the number of spin-two conserved currents.  We stress that the discussions in this section add nothing new to long-established wisdom.  We simply explain and present the results of our enumeration.

To begin, we need the basic building blocks --- the knowledge of the vacuum characters up to ${\cal O}(q^2)$ for affine $U(1)$ and all simple compact affine Lie algebras.  For $U(1)$,
\ie
c^{U(1)} = 1 \, , \quad 
\chi^{U(1)}_r(q) = 
\begin{cases}
1 + 3q + 4q^2 + {\cal O}(q^3) \quad & r = 1 \, ,
\\
1 + q + 4q^2 + {\cal O}(q^3) & r = \sqrt 2 \, ,
\\
1 + q + 2q^2 + {\cal O}(q^3) & \text{otherwise} \, .
\end{cases}
\fe
At $r = 1$, the $U(1)$ affine current algebra is enhanced to $SU(2)_1$.  At $r = \sqrt2$, it is enhanced to a $\cal W$-algebra.
For a simple compact affine Lie algebra $G_k$ ($k$ is the level), the Sugawara central charge is
\ie
c^G_k = {k \, \text{dim} \, G \over k + h^\vee} \, .
\fe
The dimension and dual coxeter number $h^\vee$ are given in Table~\ref{Tab:G}.
The character up to $q^2$ is
\ie
\chi^G_k(q) = 1 + (\text{dim} \, G) \, q + a^G_{k, 2} \, q^2 + {\cal O}(q^3) \, ,
\fe
where the $q^2$ coefficient for $k \ge 2$ is
\ie
a^G_{k, 2} = {(\text{dim} \, G) (\text{dim} \, G + 3) \over 2} \, , \quad k \ge 2 \, .
\fe
For $k = 1$, the coefficient $a^G_{1, 2}$ is given in Table~\ref{Tab:G}.\footnote{The vacuum characters of the $BD$ series at level 1 have a simple expression
\ie
\label{BDVacChar}
\chi_\text{vac}^{SO(N)_1}(\tau) &= {\theta_3(\tau)^{N\over2} + \theta_4(\tau)^{N\over2} \over 2\eta(\tau)^{N\over2}} \, ,
\fe
where $\theta_i(\tau)$ is the Jacobi theta function.
We use the Affine.m package \cite{Nazarov:2011mv} to compute $a^G_{1, 2}$ for the exceptional $EFG$ and for the $AC$ series up to $n = 6$, and deduce analytic formulae for the latter assuming a quartic ansatz.
}

\renewcommand{\arraystretch}{1.5}
\begin{table}[h]
\centering
\begin{tabular}{|c|cccc|}
\hline
$G$ & $\text{dim} \, G$ & $h^\vee$ & $c^G_1$ & 
$a^G_{1, 2} - \text{dim} \, G$
\\\hline
$U(1)$ & 1 & 0 & 1 &
\\
$A_{n\ge1}$ & $n(n+2)$ & $n+1$ & $n$ & $\displaystyle {n^2(n+1)^2\over4}$
\\
$B_{n\ge2}$ & $n(2n+1)$ & $2n-1$ & $n+{1\over2}$ & $\begin{pmatrix} 2n+2 \\ 2 \end{pmatrix} + \begin{pmatrix} 2n+1 \\ 4 \end{pmatrix}$
\\
$C_{n\ge3}$ & $n(2n+1)$ & $n+1$ & $\displaystyle {n(2n+1) \over n+2}$ & $\displaystyle {n^2 (2n+1) (2n-1) \over 3}$
\\
$D_{n\ge2}$ & $n(2n-1)$ & $2n-2$  & $n$ & $\begin{pmatrix} 2n+1 \\ 2 \end{pmatrix} + \begin{pmatrix} 2n \\ 4 \end{pmatrix}$
\\
$E_6$ & 78 & 12 & 6 
& 651
\\
$E_7$ & 133 & 18 & 7 
& 1540
\\
$E_8$ & 248 & 30 & 8 
& 3876
\\
$F_4$ & 52 & 9 & ${26 \over 5}$ 
& 325
\\
$G_2$ & 14 & 4 & ${14 \over 5}$ 
& 28
\\\hline
\end{tabular}
\caption{Data for $U(1)$ and compact simple Lie groups, including the dimension, the dual coxeter number $h^\vee$, as well as the central charge and $q^2$ coefficient of the level-1 affine current algebra.}
\label{Tab:G}
\end{table}

\begin{figure}[H]
\centering
\includegraphics[width=.475\textwidth]{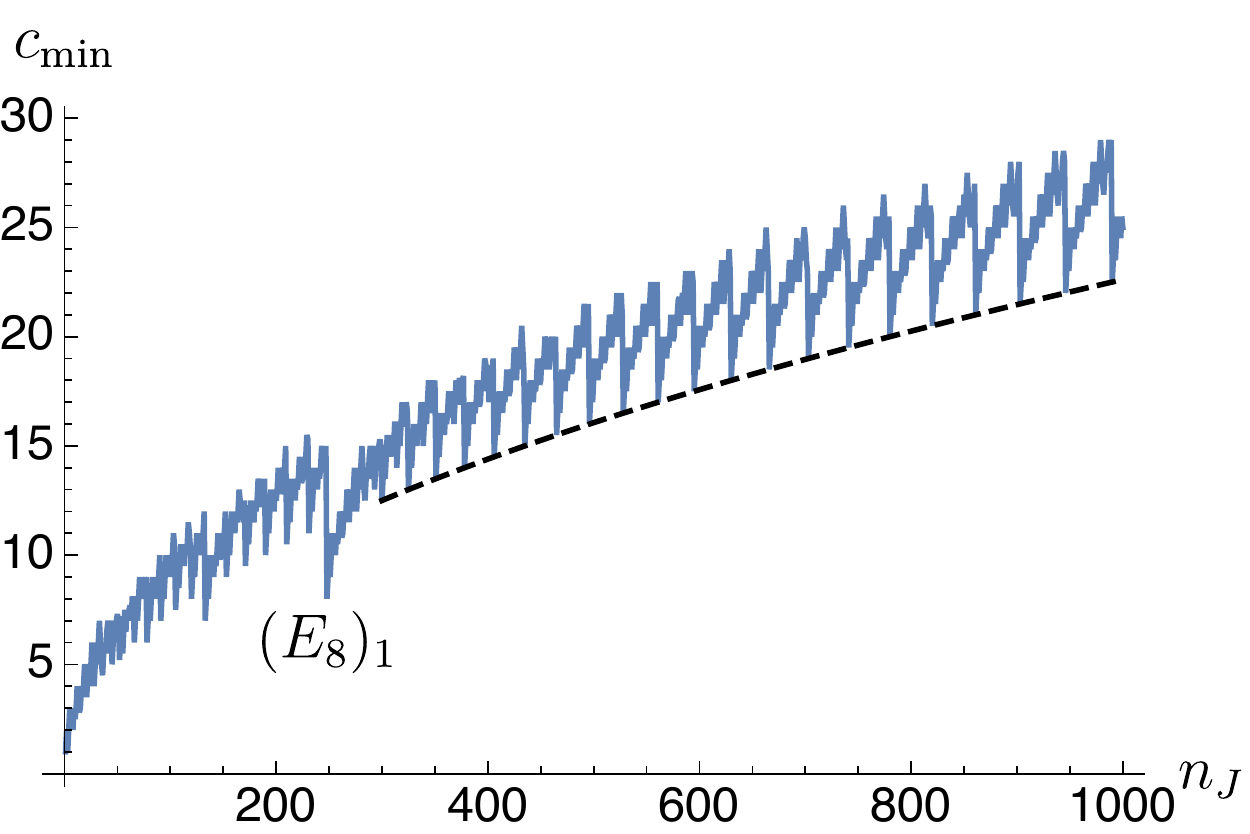}
\quad
\includegraphics[width=.475\textwidth]{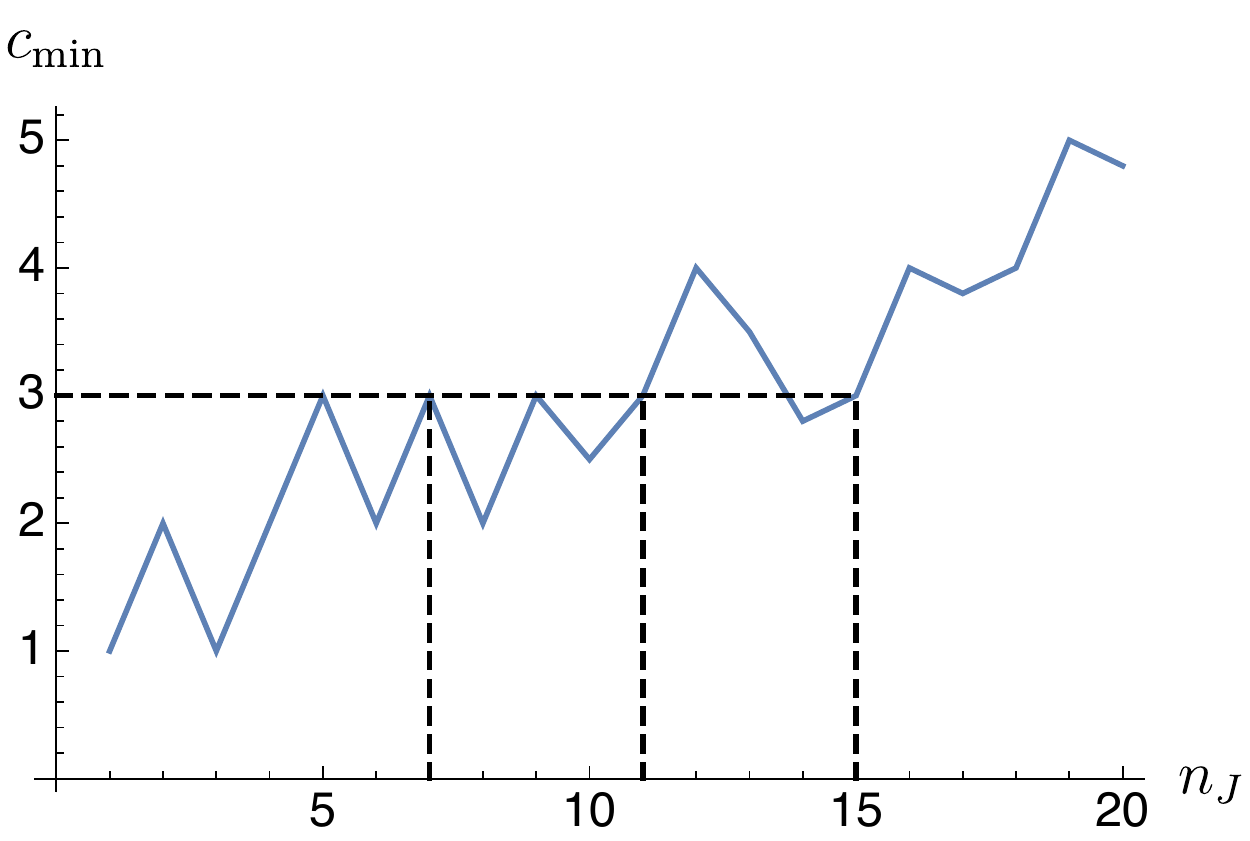}
\caption{{\bf Left:} Lower bounds on the central charge $c$ for given numbers of spin-one conserved currents $n_J$, up to $n_J = 1000$. The dashed line shows the lower envelope \eqref{Envelopes} for $n_J \ge 300$. {\bf Right:} The same bounds for $n_J \le 20$, with the dashed line indicating $n_J = 7, 11, 15$, $c_\text{min} = 3$.}
\label{Fig:cmin}
\end{figure}

\begin{figure}[H]
\centering
\includegraphics[width=.475\textwidth]{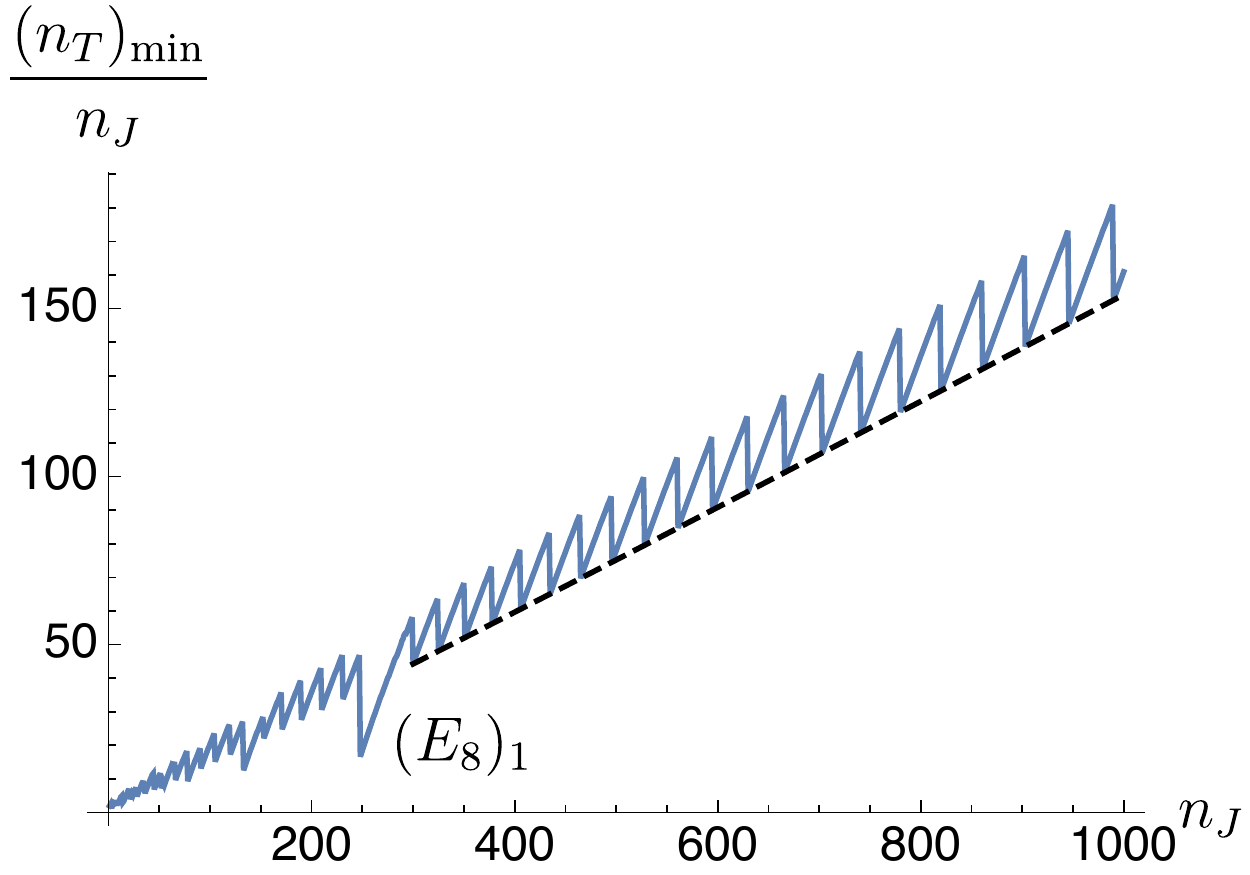}
\quad
\includegraphics[width=.475\textwidth]{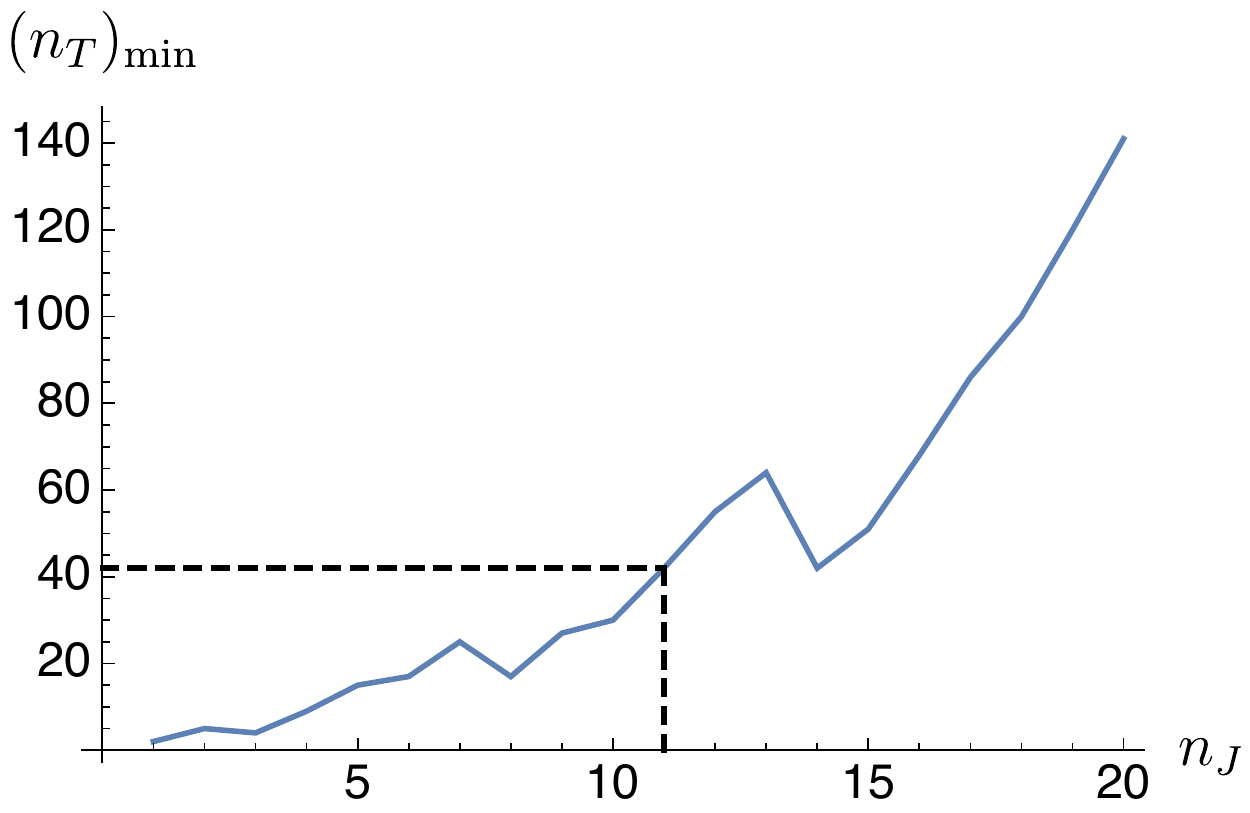}
\caption{{\bf Left:} Lower bounds on the number of spin-two conserved currents $n_T$, for given numbers of spin-one conserved currents $n_J$, up to $n_J = 1000$. The black dashed line shows the lower envelope \eqref{Envelopes} for $n_J \ge 300$.  {\bf Right:} The same bounds for $n_J \le 20$, with the dashed line indicating $n_J = 11$, $(n_T)_\text{min} = 42$.}
\label{Fig:nT}
\end{figure}

It is clear that for a fixed number $n_J$ of spin-one conserved currents, the minimum central charge $c$ and the minimum number $n_T$ of spin-two conserved currents (including primaries and descendants) are realized by products of $U(1)$ factors and compact affine simple Lie algebras \emph{at level 1}.  Moreover, since $(C_n)_1$ has the same $n_J$ as $(B_n)_1$ but strictly larger $c$ and $n_T$, we can exclude the $C$ series for the purpose of bounding $c$ and $n_T$.
The bounds up to $n_J = 1000$ are presented in Figures~
\ref{Fig:cmin} and \ref{Fig:nT}.  We find that considering products of up to three $U(1)$ or simple factors is enough to realize the minima up to $n_J = 1000$.
The teeth structure is explained as follows.  At certain $n_J$, low $c$ and $n_T$ are achieved by a {\it simple} compact affine Lie algebra, such as the $(E_8)_1$ algebra at $n_J = 248$.  The next few values of $n_J$ have minimal $c$ or $n_T$ that are simply given by their product with the affine current algebras that saturate the bounds at $n_J = 1, 2, 3, \dotsc$.  This continues until the next compact affine simple Lie algebra with especially low $c$ or $n_T$ occurs.  We find that past $(E_8)_1$, the especially small occurrences are all given by the $(B_n)_1$ and $(D_n)_1$ algebras.  Hence, when $n_J \ge 300$, the lower envelopes of the lower bounds are given by\footnote{Since the $(B_n)_1$ and $(D_n)_1$ vacuum characters have simple closed forms \eqref{BDVacChar}, one could easily compute this envelope for all higher-spin conserved currents.
}
\ie
\label{Envelopes}
& c \ge c_\text{min} = {1+\sqrt{1+8n_J}\over4} \, ,
\\
& n_T \ge (n_T)_\text{min} = {3+14n_J+n_J^2+3\sqrt{1+8n_J}-n_J\sqrt{1+8n_J} \over 6} \, .
\fe
Interesting bounds of a similar flavor were obtained in \cite{Benjamin_2016} for holomorphic CFTs.

The bound on the central charge is a quantum version of the following geometric question: What is the minimal dimension $c_\text{min}^\text{cl}$ of a compact connected manifold that realizes a continuous isometry group of dimension $n_J^\text{cl}$?  The answer is given by an analogous exercise as the above, but now the basic building blocks are the classical groups $U(1)$ and $ABD$, with\footnote{Since the dimensions of $B_n$ and $C_n$ are identical, and $B_n$ is known to be the maximal isometry group for a compact $2n$-dimensional manifold, we can ignore $C_n$ in trying to minimize $c_\text{min}^\text{cl}$.  The same cannot be said for $A_n$, which is the isometry group of $\bP^n$.
However, since the ``cost factor'' ${c_\text{min}^\text{cl} / n_J^\text{cl}}$ of $A_n$ is roughly twice of $B_n$ and $D_n$,
their exclusion turns out to not affect $c_\text{min}^\text{cl}$.
}
\ie
c_\text{min}^\text{cl} =
\begin{cases}
1 & U(1) \, ,
\\
2n & A_n \, ,
\\
2n & B_n \, ,
\\
2n-1 &  D_n \, .
\end{cases}
\fe
The especially small occurrences are spheres 
with $B_n$ and $D_n$ symmetries, giving the lower envelope
\ie
\label{ClassicalEnvelope}
c^\text{cl} \ge c_\text{min}^\text{cl} = {-1+\sqrt{1+8n_J^\text{cl}}\over2} \, .
\fe

\begin{figure}[H]
\centering
\includegraphics[width=.475\textwidth]{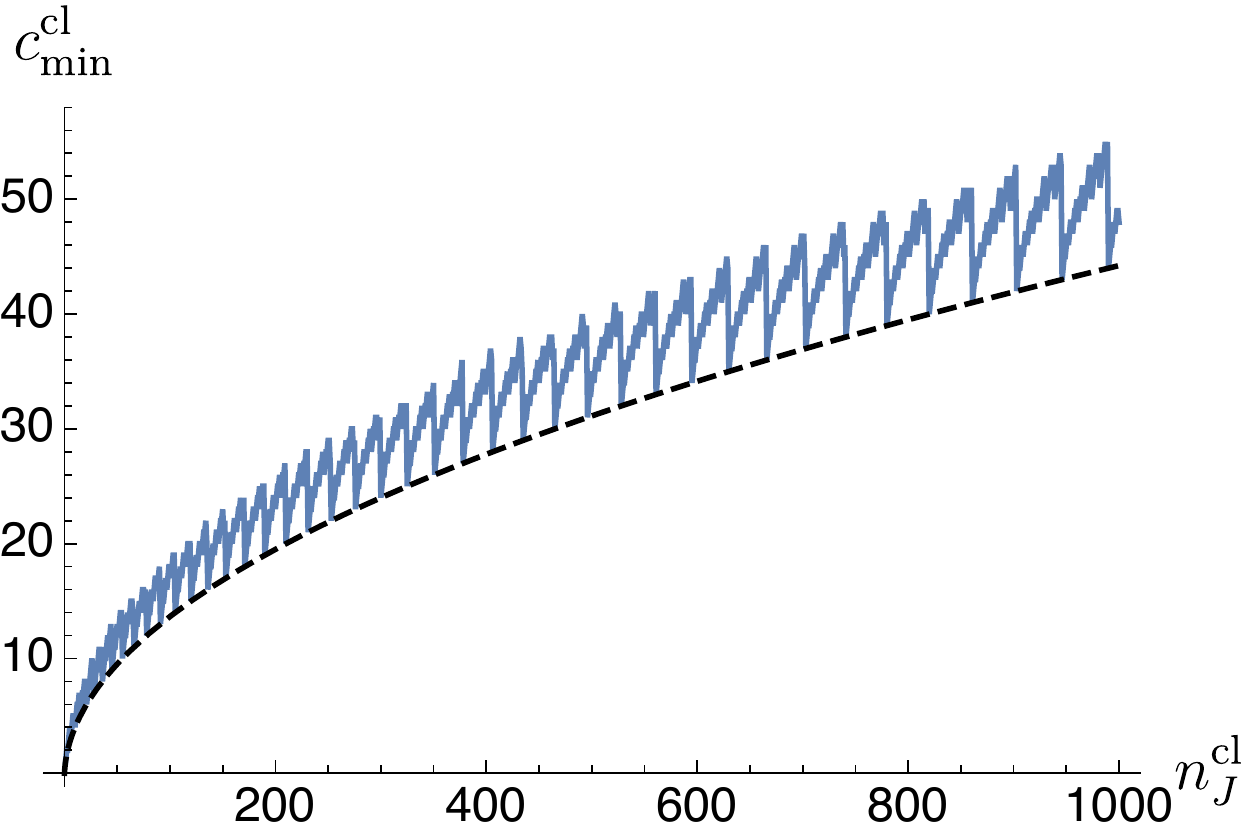}
\quad
\includegraphics[width=.475\textwidth]{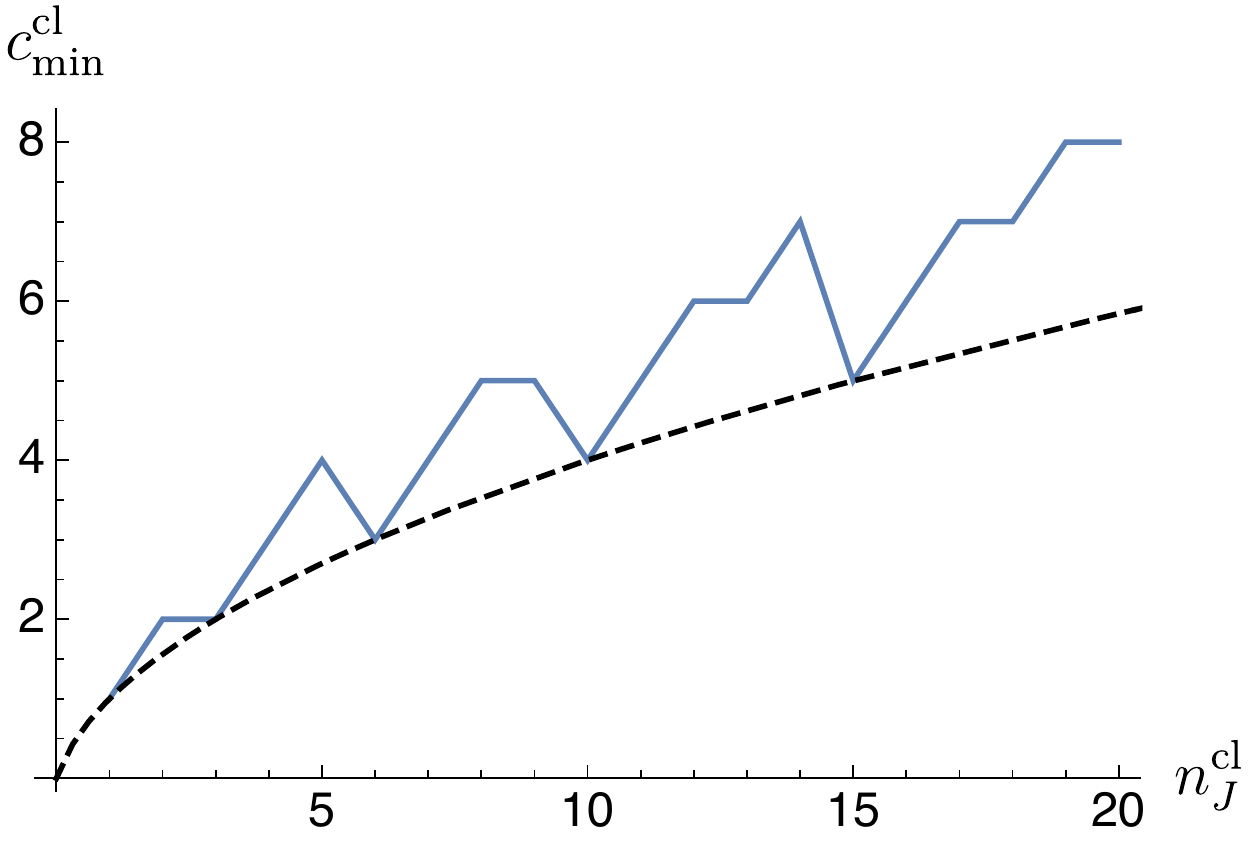}
\caption{Lower bounds on the dimension of a connected manifolds, for given dimensions of the isometry group.  The black dashed line shows the lower envelope \eqref{ClassicalEnvelope}.}
\label{Fig:Classical}
\end{figure}

\subsubsection{Patient 4: Extremal $U(1)$ flavored modular function}
\label{Sec:PatientDyer}

In \cite{Dyer:2017rul}, the modular bootstrap with a holomorphic $U(1)$ or $\bR$ flavor symmetry was studied.  For the spinning bootstrap at $c = 3$ (with half-integer spins allowed), they found an explicit form for the extremal partition function, given in (5.9) of \cite{Dyer:2017rul}, and suspected it
to be the NS+ partition function of a fermionic CFT:
\footnote{Compared to \cite{Dyer:2017rul}, we rescale $z$ and $\bar z$ by a factor of two to make all the $U(1)$ charges integers.
}
\begin{align}
Z^{\text{NS}+}(\tau, \bar\tau, z, \bar z) &= \frac{1}{4|\eta(\tau)|^2} \sum_{a, b, a', b'=0}^1 (-1)^{ab' + a'b} |\vartheta_{ab}(\tau, z)|^4 |\vartheta_{a'b'}(\tau,0)|^2
\label{eq:EthanFermionic}
\end{align}
where $\vartheta_{a,b}$ are the Jacobi theta functions.\footnote{The relation to another common notation is
\ie
\vartheta_{11}(\tau, z) = \theta_1(\tau, z), \quad \vartheta_{10}(\tau, z) = \theta_2(\tau, z), \quad \vartheta_{00}(\tau, z) =\theta_3(\tau, z), \quad \vartheta_{01}(\tau, z) = \theta_4(\tau, z) \, .
\fe
}
Assuming so, and turning off the $U(1)$ fugacity for now ($z = \bar z = 0$), the partition function of the bosonized CFT takes the form
\ie
Z^B(\tau, \bar\tau) = {1\over2} \left( Z^{\text{NS}+}(\tau,\bar \tau) + Z^{\text{NS}-}(\tau,\bar \tau) + Z^{\text{R}+}(\tau,\bar \tau) + Z^{\text{R}-}(\tau,\bar \tau) \right) \, ,
\fe
the first three of which we know by \eqref{eq:EthanFermionic} and its modular transforms.  To the lowest few orders in the $q$-expansion, they are
\ie
Z^{\text{NS}+}(\tau,\bar\tau) &= (q \bar q)^{-\frac18} \Big[ (1 + 7q + 27q^2 + \ldots) + (8q^{\frac12} + 32q^{\frac32} + \ldots)(q\bar{q})^{\frac18}  \label{eq:ethansstuffnss}
\\
&   \hspace{0.5in} + (16q^{\frac12} + 80q^{\frac32} +\ldots)(q\bar{q})^{\frac14} +  (16+96q+\ldots)(q\bar{q})^{\frac38} +\ldots \Big] 
\\
Z^{\text{NS}-}(\tau,\bar\tau) &= (q \bar q)^{-\frac18} \Big[ (1 + 7q + 27q^2 + \ldots) - (8q^{\frac12} + 32q^{\frac32} + \ldots)(q\bar{q})^{\frac18} 
\\
&   \hspace{0.5in} - (16q^{\frac12} + 80q^{\frac32} +\ldots)(q\bar{q})^{\frac14} +  (16+96q+\ldots)(q\bar{q})^{\frac38} +\ldots \Big] 
\\
Z^{\text{R}+}(\tau,\bar\tau) &= (q \bar q)^{-\frac18} \Big[ (8q + 24q^2 + \ldots) + (2 + 16q + \ldots)(q\bar{q})^{\frac18} 
\\
&   \hspace{0.5in}+  (8 + 40 q + \ldots)(q\bar{q})^{\frac14}  + (16+96q+\ldots)(q\bar{q})^{\frac38} + \ldots \Big]. 
\fe
Note an interesting feature of the NS sector: at any fixed twist, the states of that twist are either \emph{all} half-integer spin or \emph{all} integer spin.
In particular, states with $h \equiv 0, \frac38$ mod $\frac12$ all have integer spin, and states with $h \equiv \frac 18, \frac 14$ mod $\frac12$ all have half-integer spin. (Since the spin is always an integer or half-integer, we could of course replace $h$ with $\bar h$ in the previous sentence. Also there are no states with $h$ or $\bar{h} \equiv \frac 78~ \text{mod}~ 1$.)

The authors of \cite{Dyer:2017rul} asked if
\ie
\label{EthanBosonic}
\frac12\(Z^{\text{NS}+}(\tau, \bar\tau) + Z^{\text{NS}-}(\tau,\bar\tau) + Z^{\text{R}+}\(\tau, \bar\tau\)\) = (q \bar q)^{-\frac18} \( 1 + 11 q + 39 q^2 + \dotsb \) \, ,
\fe
being an $SL(2, \bZ)$ invariant function with integer spins, could be the partition function of a bosonic CFT.  Of course, physically there is no reason to expect so, since $Z^{\text{R}-}(\tau, \bar\tau)$ has no reason to vanish and should give nontrivial contributions.  Indeed, if we assume that the affine current algebra must be semi-simple and compact (possibly plus factors of $U(1)$), we can easily rule out \eqref{EthanBosonic} from the Sugawara bound.
According to Figure~\ref{Fig:cmin}, the lower bound on the central charge at $n_J = 11$ is $c \ge 3$, and is saturated by the 
affine Lie algebra $(A_1)_1 \times (A_2)_1$.  To be consistent with unitarity, 
\eqref{EthanBosonic} should be the partition function of 
the $(A_1)_1 \times (A_2)_1$ product WZW model.  But the partition function does not match,
in particular, the $(A_1)_1 \times (A_2)_1$ product WZW model has $n_T = 45 \neq 39$.  Hence \eqref{EthanBosonic} is ruled out.\footnote{Alternatively, we could rule out \eqref{EthanBosonic} by the lower bound on the number of spin-two conserved currents.  According to Figure~\ref{Fig:nT}, the lower bound  at $n_J = 11$ is $n_T \ge 42$ and violated by \eqref{EthanBosonic}. The number $n_T = 42$ is realized by the $U(1) \times (B_2)_1$ WZW model at generic $U(1)$ radius with central charge $c = {7\over2}$.
}

Next, let us examine the possibilities for $Z^{\text{R}-}(\tau, \bar\tau)$.  In the zero-twist sector, to the lowest orders in the $q$-expansion,
\ie
& Z^{\text{R}-}(\tau, \bar\tau) = (q \bar q)^{-\frac18} ( n^{\text{R}-}_1 q + n^{\text{R}-}_2 q^2 + {\cal O}(q^3) ) \, ,
\\
& n^{\text{R}-}_1 \in \{ -8, -6, \ldots, 6, 8 \} \, , \quad n^{\text{R}-}_2 \in \{ -24, -22, \ldots, 22, 24 \} \, .
\fe
Without loss of generality, we can set $n^{\text{R}-}_1 \ge 0$.
Correspondingly, the bosonized partition functions $\cal B$ and ${\cal B}/\bZ_2$ have zero-twist sectors
\ie
Z^{\cal B} &= (q \bar q)^{-\frac18} \( 1 + (11 + {n^{\text{R}-}_1 \over 2}) q + (39 + {n^{\text{R}-}_2 \over 2}) q^2 + {\cal O}(q^3) \) \, ,
\\
Z^{{\cal B}/\bZ_2} &= (q \bar q)^{-\frac18} \( 1 + (11 - {n^{\text{R}-}_1 \over 2}) q + (39 - {n^{\text{R}-}_2 \over 2}) q^2 + {\cal O}(q^3) \) \, ,
\fe
and {\it both} must satisfy the Sugawara bounds at $c = 3$.  We have already ruled out $n_1^{\text{R}-} = 0$.  From Figure~\ref{Fig:cmin}, we see that $n^{\text{R}-}_1 = 2, \, 4$ are not allowed because $\cal B$ will have $n_J = 12, \, 13$ and thus necessarily $c > 3$.  If $n^{\text{R}-}_1 = 6$, then $n_J = 14$ requires $\cal B$ to contain the $(G_2)_1$ affine current algebra, but since the Sugawara central charge is already $14\over5$, the coset part can only have central charge $1\over5$ and is in tension with unitarity.

The only possibility left is $n_1^{\text{R}-} = 8, \, n^{\text{R}-}_2 = 24$, which corresponds to ${\cal B} = (A_3)_1$ WZW model (free boson on the $A_3$ lattice), and ${\cal B}/\bZ_2 = (A_1)_1 \times (A_1)_1 \times U(1)_{r = \sqrt2}$ WZW (free boson on a rectangular lattice with radii $1, \, 1, \, \sqrt2$). This possibility is only allowed if the average of the partition functions of the $(A_3)_1$ and the $(A_1)_1 \times (A_1)_1 \times U(1)_{r=\sqrt 2}$ WZW models is equal to \eqref{EthanBosonic}. Remarkably, we find that it is.\footnote{We used the Affine.m package \cite{Nazarov:2011mv} to compute the $(A_3)_1$ characters to sufficiently high $q$ order.}
This determines $Z^{\text{R}-}(\tau, \bar\tau)$ to be
\ie
Z^{\text{R}-}(\tau,\bar\tau) &= Z^{\text{NS}+}(\tau, \bar\tau) + Z^{\text{NS}-}(\tau,\bar\tau) + Z^{\text{R}+}\(\tau, \bar\tau\) - 2Z^{(A_1)_1 \times (A_1)_1 \times U(1)_{r = \sqrt2}}(\tau,\bar\tau) \\
&= 2Z^{(A_3)_1}(\tau,\bar\tau) - (Z^{\text{NS}+}(\tau, \bar\tau) + Z^{\text{NS}-}(\tau,\bar\tau) + Z^{\text{R}+}\(\tau, \bar\tau\) ) \\
&= (q \bar q)^{-\frac18} \Big[ (8q + 24q^2 + \ldots) - (2 + 16q + \ldots)(q\bar{q})^{\frac18} \label{eq:zrminusethan}
\\& \hspace{0.5in} - (8 + 40 q + \ldots)(q\bar{q})^{\frac14}  + (16+96q+\ldots)(q\bar{q})^{\frac38} + \ldots \Big].
\fe
In (\ref{eq:ethansstuffnss}) and (\ref{eq:zrminusethan}), we find that the fermion parity $(-1)^F$ in both the NS and Ramond sectors follow a simple pattern: if $h \equiv 0, \frac 38$ mod $\frac12$, the state is bosonic; if $h \equiv \frac18, \frac14$ mod $\frac12$, the state is fermionic. 
Using the isomorphism of Hilbert spaces in Table~\ref{Fig:Square}, the previous observation allows us to identify by which $\bZ_2$ global symmetry to fermionize ${\cal B}/\bZ_2$:  It is the product of the $\bZ_2$ center symmetry of both copies of $(A_1)_1$, and the momentum $\bZ_2$ symmetry of $U(1)_{r = \sqrt2}$.\footnote{Note that the $\bZ_2$ center symmetry in a single copy of $(A_1)_1$ is anomalous, but the product of two of them is non-anomalous.
}

We can also identify the $U(1)$ symmetry by which \eqref{eq:EthanFermionic} is graded.  In the bosonized theory $\mathcal{B}/\mathbb Z_2$, the holomorphic part is the momentum plus winding $U(1)$ in one copy of $SU(2)_1$, while the anti-holomorphic part is the momentum minus winding $U(1)$ in the other copy of $SU(2)_1$.  After fermionizing by the $\mathbb{Z}_2$ described in the previous paragraph, we successfully reproduce the graded extremal partition function in \cite{Dyer:2017rul}.

\section{RNS modular bootstrap}
\label{sec:lessons}

The modular bootstrap has been applied to extended ${\cal N} = 2$ supersymmetry in \cite{Friedan:2013cba,Keller:2012mr} and to ${\cal N} = 1$ and non-extended ${\cal N} = 2$ in \cite{Bae:2018qym}.\footnote{Extended ${\cal N} = 2$ has spectral flow symmetry, with the level of the $U(1)_R$ $R$-symmetry given by $c/3$.
}
In those work, only the modular invariance of the NS sector was studied.  With extended ${\cal N} = 2$ supersymmetry, this is because the Ramond sector is isomorphic to the NS sector by a half-integer spectral flow.  However, when spectral flow is not a symmetry, the consistency of the Ramond sector is not guaranteed by the $\Gamma_\theta$ invariance of the NS+ partition function.

In this section, we strengthen the ${\cal N} = 1$ modular bootstrap analysis of \cite{Bae:2018qym} by incorporating the Ramond sector.
Indeed, some putative partition functions of \cite{Bae:2018qym} are ruled out by our bounds (and in fact ruled out by the simple diagnostics of Section~\ref{Sec:Diagnostic}).  The bounds on the scalar gap also prove that relevant deformations must exist in certain ranges of the central charge.  The ranges are different depending on how much manifest supersymmetry we preserve.\footnote{The analogous study in \cite{Bae:2018qym} did not consider supersymmetry-preserving relevant {\it descendants}.  We explain this point in Section~\ref{Sec:Bootstrap}.
}

What about the non-supersymmetric modular bootstrap of fermionic CFTs?  Via the bosonization/fermionization map and the isomorphism of Hilbert spaces shown in Figure~\ref{Fig:Square}, non-supersymmetric fermionic modular bootstrap is in fact identical to the bosonic modular bootstrap with a non-anomalous $\bZ_2$ global symmetry.  This was previously studied in \cite{Lin:2019kpn} by one of the present authors.

\subsection{${\cal N} = 1$ characters}

We assume $c>{3\over2}$ so that 
the possible ${\cal N} = 1$ super-Virasoro modules are universal. The characters are given by  
\begin{align}
\chi_h^{\text{NS}}(\tau) &= \begin{cases}
\displaystyle q^{-\frac{c-\frac32}{24}}(1-q^{\frac12})\frac{\eta(\tau)}{\eta(2\tau)\eta(\tau/2)} \qquad & h = 0
\\
\displaystyle q^{h-\frac{c-\frac32}{24}} \frac{\eta(\tau)}{\eta(2\tau)\eta(\tau/2)} \qquad & h > 0
\end{cases}
\end{align}
in the NS sector and
\begin{align}
\chi_h^{\text{R}}(\tau) = \begin{cases} \displaystyle \frac{\eta(2\tau)}{\eta(\tau)^2} & h = \frac{c}{24} 
\\ 
\displaystyle 2q^{h-\frac{c}{24}} \frac{\eta(2\tau)}{\eta(\tau)^2} \qquad & h > \frac c{24} \end{cases}
\label{eq:Rchar}
\end{align}
in the Ramond sector\footnote{The factor of 2 for non-BPS characters in (\ref{eq:Rchar}) is due to the fact that $G_0$ does not annihilate the highest weight state for non-BPS primaries.}.  
Given the mathematical identities
\ie
\frac{\eta(\tau)}{\eta(2\tau)\eta(\tau/2)} &= \frac{\eta(-1/\tau)}{\eta(-2/\tau)\eta(-1/2\tau)} \times (-i\tau)^{1\over2} \, ,
\\
\frac{\eta(\tau)}{\eta(2\tau)\eta((\tau+1)/2)} &= \frac{\eta(-2/\tau)}{\eta(-1/\tau)^2} \times e^{-{\pi i \over24}} (-2i\tau)^{1\over2} \, ,
\fe
if we are only concerned with the $S$ transform, we could alternatively consider reduced partition functions and characters defined in Appendix~\ref{App:Reduced}.

The torus partition functions 
are each given by a sum over ${\cal N} = 1$ characters
\begin{align}
Z^{\text{R}+}(\tau,\bar\tau)&=\sum_{(h,\bar h)\in {\cal H}_\text{R}} n^\text{R}_{h,\bar h}\, \chi_h^\text{R}(\tau)\chi_{\bar h}^\text{R}(\bar\tau)\,,
\\
Z^{\text{NS}-}(\tau,\bar \tau)&=\sum_{(h,\bar h)\in {\cal H}_\text{NS}} n^\text{NS}_{h,\bar h}\, \chi_h^\text{NS}(\tau+1) \chi_{\bar h}^\text{NS}(\bar\tau+1)\,,
\\
Z^{\text{NS}+}(\tau,\bar \tau)&=\sum_{(h,\bar h)\in {\cal H}_\text{NS}} n^\text{NS}_{h,\bar h}\, \chi_h^\text{NS}(\tau) \chi_{\bar h}^\text{NS}(\bar\tau)
\label{ZL}
\end{align}
with non-negative integer degeneracies $n_{h, \bar h}$.
The Witten index $Z^{\text{R}-}(\tau, \bar\tau) = \chi$ almost decouples from the modular crossing equation except that it imposes a bound $n^\text{R}_{{c\over24}, {c\over24}} \ge |\chi|$.  Its sign can be changed by tensoring with the fermionic SPT $(-1)^\text{Arf}$, as explained in Section~\ref{sec:bosonizationbohnanza}.

\subsection{Modular crossing equation}
\label{Sec:Bootstrap}

The crossing equations for the torus partition functions under the modular $S$ transform are simply
\ie
Z^{\text{NS}+}(\tau, \bar\tau) &= Z^{\text{NS}+}\left(-\frac1\tau, -\frac1{\bar\tau}\right) \, ,
\quad
Z^{\text{NS}-}(\tau, \bar\tau) = Z^\text{R}\left(-\frac1\tau, -\frac1{\bar\tau}\right) \, .
\fe
Expanded in characters,
\ie
0 &= \sum_{(h,\bar h)\in {\cal H}_\text{NS}} n^\text{NS}_{h,\bar h}\,
\begin{pmatrix}
\chi_h^\text{NS}\left(\tau\right) \chi_{\bar h}^\text{NS}\left({\bar\tau}\right) - \chi_h^\text{NS}\left(-1/\tau\right) \chi_{\bar h}^\text{NS}\left(-1/{\bar\tau}\right)
\\
\chi_h^\text{NS}\left(\tau+1\right) \chi_{\bar h}^\text{NS}\left({\bar\tau}+1\right)
\end{pmatrix} \,
\\
& \qquad + \sum_{(h,\bar h)\in {\cal H}_\text{R}} n^\text{R}_{h,\bar h}\, 
\begin{pmatrix}
0
\\
- \chi_h^\text{R}(-1/\tau)\chi_{\bar h}^\text{R}(-1/\bar\tau)
\end{pmatrix}
\, ,
\fe
where the ground state contribution is
\ie
\begin{pmatrix}
\chi_0^\text{NS}\left(\tau\right) \chi_0^\text{NS}\left({\bar\tau}\right) - \chi_0^\text{NS}\left(-1/\tau\right) \chi_0^\text{NS}\left(-1/{\bar\tau}\right)
\\
\chi_0^\text{NS}\left(\tau+1\right) \chi_0^\text{NS}\left({\bar\tau}+1\right)
\end{pmatrix}
+
n^\text{R}_{{c\over24},{c\over24}}\, 
\begin{pmatrix}
0
\\
- \chi_{c\over24}^\text{R}(-1/\tau)\chi_{c\over24}^\text{R}(-1/\bar\tau)
\end{pmatrix} \, .
\fe
For simplicity, in the following we write $n^\text{R}_{{c\over24},{c\over24}}$ as $n^\text{R}_\text{gs}$.
Note that because the ground state character is a smooth limit of the non-degenerate character in the Ramond sector, $n^\text{R}_\text{gs}$ is only a lower bound on the ground state degeneracy, unless a gap in the Ramond sector is imposed.

We use the linear functional method \cite{Rattazzi:2008pe} to analyze the modular crossing equation.  The computations are performed with the semidefinite programming solver SDPB2 \cite{Simmons-Duffin:2015qma, Landry:2019qug}.  The reader is referred to \cite{Lin:2019kpn} by one of the present authors for the details of the numerical implementation.  We simply note that the bounds presented here are at derivative order $\Lambda = 19$.  Since this method produces bounds on various types of quantities, we will first decide on a physically interesting quantities to study.

\subsection{Bounds on relevant deformations}

In the ${\cal N} = 1$ context, Lorentz-preserving operator deformations come in four classes that preserve different amounts of supersymmetry, and are listed in Table~\ref{Tab:DeformationOperators}.
Descendants involving super-Virasoro generators outside those in Table~\ref{Tab:DeformationOperators} are total derivatives, and give rise to trivial deformations.  A relevant deformation exists if there exists a scalar operator of scaling dimension less than 2.
As the modular bootstrap is adept at bounding scaling dimensions, we will be able to show that there is a critical $c$ below which a relevant deformation must exist.  Our analysis extends the work of \cite{Collier:2016cls} in the bosonic setting.

\begin{table}[H]
\renewcommand{\arraystretch}{1.5}
\centering
\begin{tabular}{|c|c|}
\hline
Scalar operator & Preserved supersymmetry
\\\hline
$\phi$ & ${\cal N} = (0,0)$
\\
$\bar G_{-{1\over2}} \cdot \psi_{s = {1\over2}}$ & ${\cal N} = (0,1)$
\\
$G_{-{1\over2}} \cdot \psi_{s = -{1\over2}}$ & ${\cal N} = (1,0)$
\\
$G_{-{1\over2}} \bar G_{-{1\over2}} \cdot \phi$ & ${\cal N} = (1,1)$
\\\hline
\end{tabular}
\caption{Operator deformations in an ${\cal N} = (1,1)$ SCFT.  Here, $\phi$ denotes a bosonic super-Virasoro primary, and $\psi$ denotes a fermionic one, both in the NS sector.}
\label{Tab:DeformationOperators}
\end{table}

\begin{figure}[H]
\centering
\includegraphics[width=.55\textwidth]{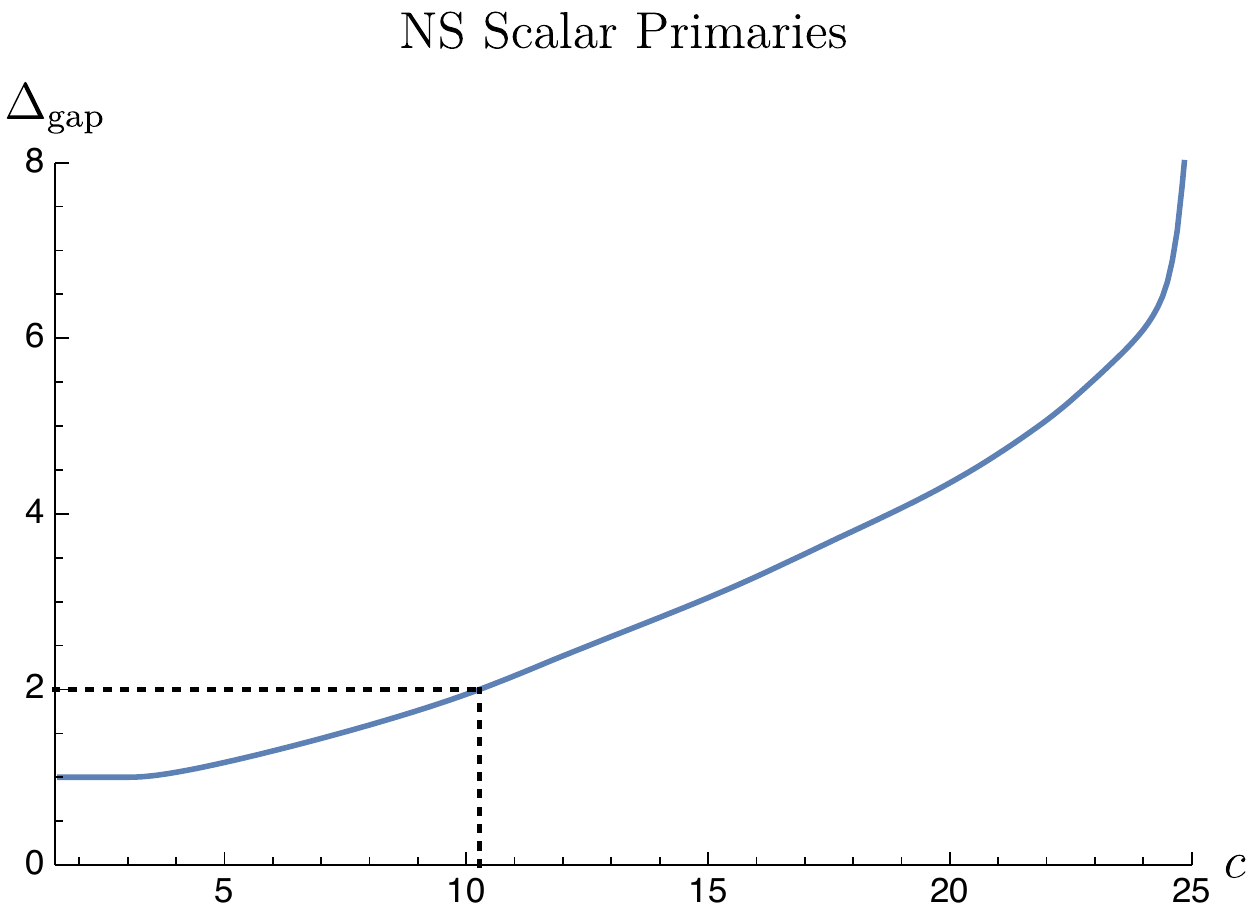}
\caption{Upper bounds on the gap in the NS spectrum of scalar super-Virasoro primaries, as the central charge is varied up to $c=25$.  The range of $c$ in which the bound is below 2, marked by the dashed lines, indicates the necessary existence of a supersymmetry breaking relevant deformation.}
\label{Fig:1}
\end{figure}

\begin{figure}[H]
\centering
\includegraphics[width=.55\textwidth]{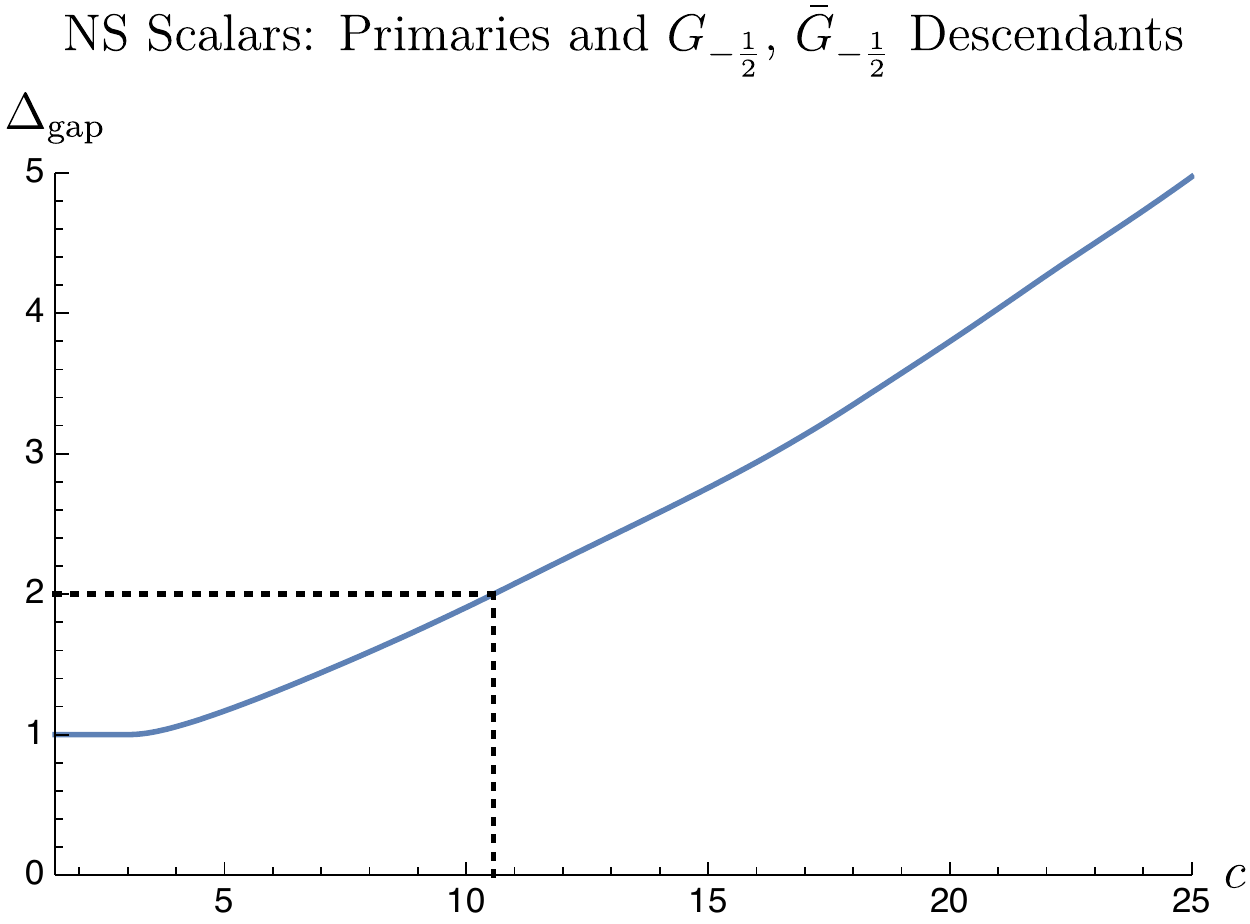}
\caption{Upper bounds on the gap in the NS spectrum of scalar operators that are either super-Virasoro primaries or $G_{-{1\over2}}$, $\bar G_{-{1\over2}}$ descendants (of fermionic superconformal primaries), as the central charge is varied up to $c=25$.  The range of $c$ in which the bound is below 2, marked by the dashed lines, indicate the necessary existence of a relevant deformation.}
\label{Fig:2}
\end{figure}

\begin{figure}[H]
\centering
\includegraphics[width=.55\textwidth]{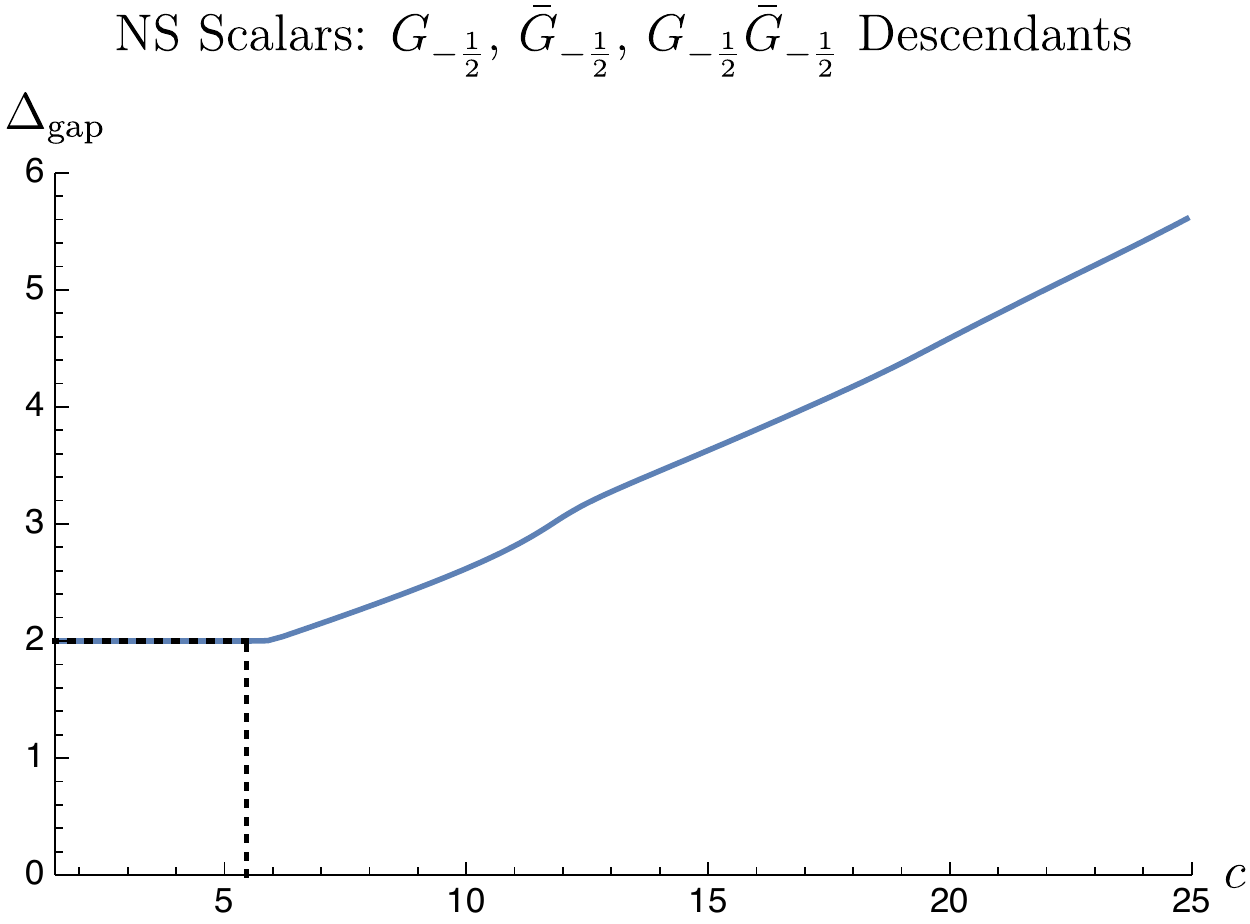}
\caption{Upper bounds on the gap in the NS spectrum of scalar operators that are $G_{-{1\over2}}$, $\bar G_{-{1\over2}}$ or $G_{-{1\over2}} \bar G_{-{1\over2}}$ descendants, as the central charge is varied up to $c=25$.  The range of $c$ in which the bound is at 2, marked by the dashed lines, indicate the necessary existence of a marginal or relevant deformation that preserved some supersymmetry.}
\label{Fig:7}
\end{figure}

\begin{figure}[H]
\centering
\includegraphics[width=.55\textwidth]{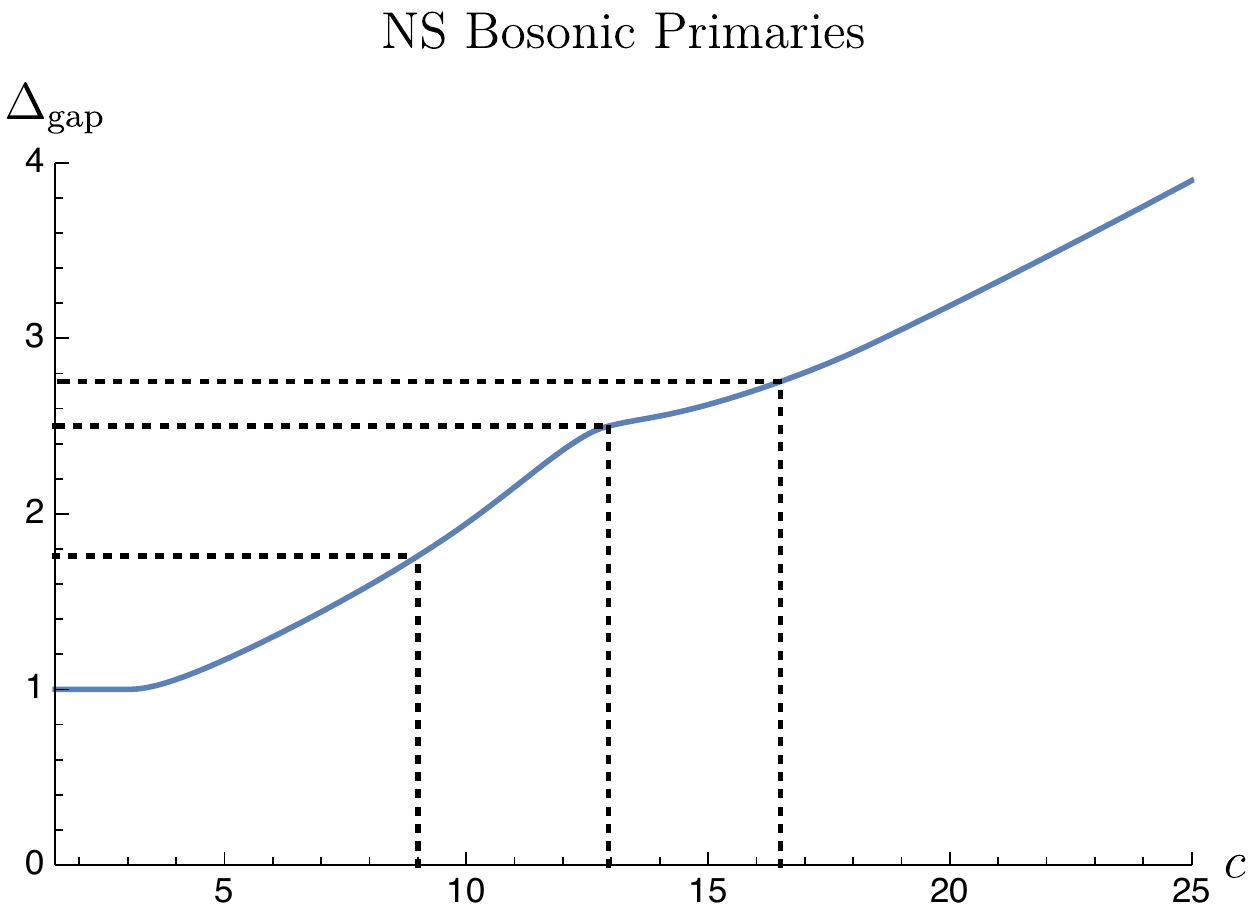}
\caption{Upper bounds on the gap in the NS bosonic primary spectrum, as the central charge is varied up to $c=25$. The dashed lines mark the bounds at $c = 9, 13, {33\over2}$.}
\label{Fig:3}
\end{figure}

\begin{figure}[H]
\centering
\includegraphics[width=.7\textwidth]{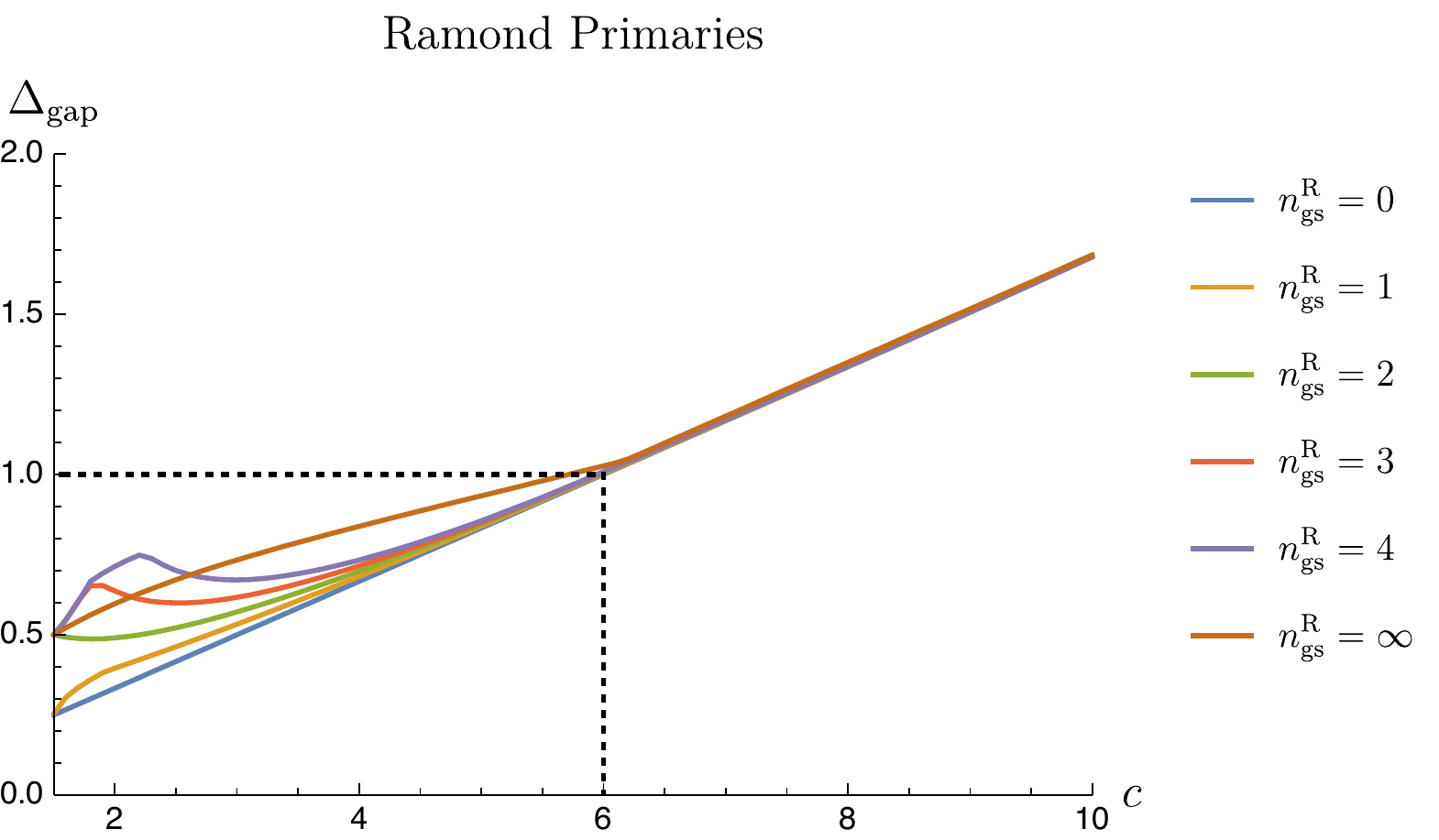}
\caption{Upper bounds on the gap in the Ramond primary spectrum, for varying number of ground states $n^\text{R}_\text{gs} = 0, 1, 2, 3, 4, \infty$, and as the central charge is varied up to $c=10$.
}
\label{Fig:4}
\end{figure}

We study the upper bound on the gap in the following subsectors of the spectrum, as the central charge is varied:
\begin{enumerate}
\item Scalar super-Virasoro primaries in the NS sector.  The results are presented in Figure~\ref{Fig:1}.  A bound below 2 means that there must exist supersymmetry-{\it breaking} relevant deformations.  We find that this is the case for
\ie
{3\over2} < c < 10.3 \, .
\fe
This was studied in \cite{Bae:2018qym} by the $\Gamma_\theta$ invariance of $Z^{\text{NS}+}$ alone, which produced a significantly smaller range of $c$.
\item Scalar operators in the NS sector that are super-Virasoro primaries or $G_{-{1\over2}}$, $\bar G_{-{1\over2}}$ descendants.  The results are presented in Figure~\ref{Fig:2}.  A bound below 2 means that there must exist relevant deformations, regardless of the amount of supersymmetry preserved.  We find that this is the case for
\ie
{3\over2} < c < 10.6 \, .
\fe
\item Scalar operators in the NS sector that are $G_{-{1\over2}}$, $\bar G_{-{1\over2}}$ or $G_{-{1\over2}} \bar G_{-{1\over2}}$ descendants.  The results are presented in Figure~\ref{Fig:7}.  A bound at 2 means that there must exist relevant \emph{or marginal} deformations that preserve {\it some} supersymmetry.  We find that this is the case for
\ie
{3\over2} < c < 5.5 \, .
\fe
\item Scalar operators in the NS sector that are $G_{-{1\over2}} \bar G_{-{1\over2}}$ descendants.  The bound is above 2 for all $c > {3\over2}$.
\item Bosonic super-Virasoro primaries in the NS sector.  The results are presented in Figure~\ref{Fig:3}.  There is a soft kink near $(c, \Delta_\text{gap}) \approx (13, 2.5)$, which upon further examination does not seem to be physical.  The NS+ partition functions at $c=9$ and $c={33\over2}$ found in \cite{Bae:2018qym} violate our bounds.
\item Super-Virasoro primaries in the Ramond sector, with the number of Ramond sector grounds states varied among $n^\text{R}_\text{gs} = 0, 1, 2, 3, 4, \infty$.  The results are presented in Figure~\ref{Fig:4}.  The nontrivial dependence on $n^\text{R}_\text{gs}$ can be seen at small $c$, but weakens at larger values of $c$. For $n^\text{R}_\text{gs} = \infty$, there appears to be an anti-kink at $(c, \Delta_\text{gap}) \approx (6, 1)$, but we do not know its interpretation.
\end{enumerate}

\section{Conclusion and discussion}
\label{sec:conclude}

This work examined new consistency conditions on the partition functions of (1+1)$d$ fermionic CFTs.  While past focus \cite{Keller:2012mr, Friedan:2013cba, Bae:2018qym} has been on the $\Gamma_\theta$ invariance of the NS+ partition function, we showed that the consistency of the Ramond sector imposes nontrivial constraints.  In some examples, we found the inconsistency manifest in the non-integrality of the degeneracies or the violation of the ${\cal N} = 1$ unitarity bounds.  In one example \cite{Dyer:2017rul}, we were able to uniquely identify the previously unknown fermionic CFT from the NS+ partition function, by considering the Sugawara consistency of its bosonization and the bosonization of its tensor with the fermionic SPT $(-1)^\text{Arf}$ (see Figure~\ref{Fig:Square}).  We ended with a systematic numerical study of modular constraints on all partition functions in ${\cal N} = (1,1)$ CFT, and obtained bounds much stronger than from the $\Gamma_\theta$ invariance of the NS+ partition function alone \cite{Bae:2018qym}.  We found that a relevant deformation must exist for ${3\over2} < c < 10.6$, and one that manifestly breaks all supersymmetry must exists for ${3\over2} < c < 10.3$.

For the ${\cal N} = 1$ Maloney-Witten partition function \cite{Maloney:2007ud}, we considered the inclusion of the modular sums of more general characters, and found that appropriate combinations can save the original Maloney-Witten partition function from negative degeneracies/density of primaries.  
Holographically, the Ramond ground state degeneracy corresponds to the quantum entropy of a zero-mass BTZ black hole with periodic spin structure at zero temperature. Suppose in a certain setup, one could do the microstate counting, then one may desire an ${\cal N} = 1$ Maloney-Witten partition function that approximates the true partition function by having the correct Ramond ground state degeneracy.  As another example, in a chaotic quantum theory without extra global symmetries, there is typically no degeneracy of states. It may then be natural to expect the Ramond ground state degeneracy to be zero.  According to Table~\ref{tab:rsector}, we can adjust the Ramond ground state degeneracy and the NS degeneracy at $h = \bar h = {c-{3\over2} \over 24}$, by adding to the original Maloney-Witten partition function ($\Gamma^0(2)$ sum of NS vacuum) an appropriate linear combination of the $c = {3\over2}$ free theory partition function, the $\Gamma^0(2)$ sum of low twist NS sector scalar primaries, and the $\Gamma_0(2)$ sum of Ramond sector scalar primaries.

The constraints from the Sugawara construction is interesting from the modular bootstrap point of view.  It is known that the linear functional approach to modular bootstrap does not produce an upper bound on the number of conserved currents without further assumptions on the spectrum beyond unitarity.  The Sugawara construction does give such an upper bound for fixed values of $c$, as is clear by turning Figure~\ref{Fig:cmin} sideways.  It would be interesting to see whether incorporating the Sugawara bound into the modular bootstrap gives stronger constraints.
Morally, the Sugawara bounds
show that a certain amount of degrees of freedom (central charge $c$) is necessary to accommodate a certain number of spin-one conserved currents.  From \eqref{Envelopes}, we see that\footnote{Bounds of a similar flavor were obtained in \cite{Lin2019BoundsOT,Alavirad2019AnomaliesAU} on a related question: How many (charged) degrees of freedom are necessary to accommodate a certain amount of ’t Hooft anomaly?  In the 4$d$ ${\cal N} = 2$ context, the Sugawara construction in the 2$d$ chiral algebra together with 4$d$ unitarity imposes interesting bounds on the central charge and flavor central charge\cite{Beem_2015,Lemos:2015orc,Beem:2018duj}.
}
\ie
{c \over \sqrt{n_J}} \ge {1\over\sqrt2} \, .
\fe
Could there be generalizations of this to higher-spin conserved currents?  On the one hand, the lack of a systematic classification of $\cal W$-algebras makes it difficult to prove any such bound.\footnote{We thank Eric Perlmutter for discussions.
}
On the other hand, there exists a nontrivial lower bound on the central charge by unitarity for every known $\cal W$-algebra.  After all, we do not expect there to be a unitary CFT with $c < 1 + 10^{-10}$ and $10^{10}$ spin-two conserved currents.  Do you?

\section*{Acknowledgements}

We are grateful to Chi-Ming Chang, Scott Collier, Alex Maloney, Hirosi Ooguri, Eric Perlmutter, Shu-Heng Shao, David Simmons-Duffin, and Yifan Wang for interesting discussions, and to Jin-Beom Bae, Liam Fitzpatrick, Sungjay Lee, Shu-Heng Shao, and Jaewon Song for helpful comments on the draft. We thank the hospitality of the Bootstrap 2019 Conference at the Perimeter Institute for Theoretical Physics.  NB is supported in part by the Simons Foundation Grant No. 488653. YL is supported by the Sherman Fairchild Foundation, and by the U.S. Department of Energy, Office of Science, Office of High Energy Physics, under Award Number DE-SC0011632.

\appendix

\section{Partition function of the kink at $c={33\over2}$}
\label{sec:jaewon332}

Here we summarize the partition function of the kink found at $c={33\over2}$ in \cite{Bae:2018qym}. Consider (A.8) in \cite{Bae:2018qym} reproduced below:
\be
\left[\mathcal{D}_\tau^3 + \mu_1 M_2(\tau) \mathcal D_\tau^2 + \mu_2 M_4^{(1)}(\tau) \mathcal{D}_\tau + \mu_3 M_4^{(2)}(\tau) \mathcal{D}_\tau + \mu_4 M_6^{(1)}(\tau) + \mu_5 M_6^{(2)}(\tau) \right] f(\tau) = 0,
\label{eq:mde}
\ee
where\footnote{In (\ref{eq:asdf}), we have corrected a very minor typo in (A.2) of \cite{Bae:2018qym}.}
\be
\mathcal{D}_\tau^n \equiv \mathcal{D}_{\tau, 2n-2} \circ \mathcal{D}_{\tau, 2n-4} \circ \ldots \circ \mathcal{D}_{\tau, 2} \circ \mathcal{D}_{\tau,0}
\label{eq:asdf}
\ee
and 
\begin{align}
\mathcal{D}_{\tau,r} &\equiv \frac{1}{2\pi i}\(\partial_\tau - \frac{i\pi r}6 E_2(q)\) = q\partial_q - \frac{r}{12} E_2(q) \nn\\
E_2(q)&= 1- 24\sum_{n=1}^{\infty} \sigma_1(n) q^n, ~~~~~\sigma_x(n) = \sum_{d|n} d^x.
\label{eq:asdfg}
\end{align}
The $M_i(\tau)$ are modular forms of weight $i$ under $\Gamma_{\theta}$, and are given explicitly in (A.4) and (A.5) of \cite{Bae:2018qym}. Equation (\ref{eq:mde}) is a general modular differential equation of third order; for our purposes we will need specific choices of the $\mu$. In order to find the $\mu$ needed to match the partition function at $c={33\over2}$, it is enough to plug in an ansatz for $f(\tau)$ at $c=\frac{33}2$ with zero states at dimension ${1\over2}$. This will give the most general solution of $\mu$ as
\begin{align}
\mu_2 &= - \frac{84785+45264\mu_1}{281088} \nn\\
\mu_3 &= -\frac{72077+21360\mu_1}{281088} \nn\\
\mu_4 &= -\frac{55(349+2032\mu_1)}{499712} \nn\\
\mu_5 &= -\frac{11(-11549+23920\mu_1)}{499712}.
\end{align}
For simplicity we set $\mu_1=0$ to get
\begin{align}
\mu_2 &= -\frac{84785}{281088}\nn\\
\mu_3 &= -\frac{72077}{281088}\nn\\
\mu_4 &= -\frac{19195}{499712} \nn\\
\mu_5 &= \frac{127039}{499712}.
\label{eq:mu2345}
\end{align}
We can then plug (\ref{eq:mu2345}) into (\ref{eq:mde}) and compute $f^{c={33\over2}}(\tau)$ to arbitrarily high order.

\section{$c=\frac32$ free theory partition function}
\label{sec:c32ft}

The partition function of a $c=\frac32$ theory given by the tensor product of a free boson (at radius $r$) and a free fermion is: 
\be
Z^\text{NS}(\tau,\bar\tau) = \left|\frac{\eta(\tau)}{\eta(\tau/2)\eta(2\tau)}\right|^2\sum_{n,m\in \mathbb Z}q^{\frac14\left(\frac nr + mr\right)^2} \bar{q}^{\frac14\left(\frac nr - mr\right)^2},
\ee
\be
Z^{\text{NS}-}(\tau,\bar\tau) = \left|\frac{\eta(\tau/2)}{\eta(\tau)^2}\right|^2\sum_{n,m\in \mathbb Z}q^{\frac14\left(\frac nr + mr\right)^2} \bar{q}^{\frac14\left(\frac nr - mr\right)^2},
\ee
\be
Z^{\text{R}+}(\tau,\bar\tau) = \left|\frac{\sqrt2\eta(2\tau)}{\eta(\tau)^2}\right|^2\sum_{n,m\in \mathbb Z}q^{\frac14\left(\frac nr + mr\right)^2} \bar{q}^{\frac14\left(\frac nr - mr\right)^2},
\ee
and finally
\be
Z^{\text{R}-}(\tau,\bar\tau) = 0.
\ee
This has two R-sector ground states, and zero Witten index.

The scaling gap to the first NS primary depends on the radius $r$: $\text{min}(\frac1{2r^2}, \frac{r^2}2)$. The Ramond sector gap is similarly $\text{min}(\frac1{2r^2}, \frac{r^2}2)$.

We can see that the modular crossing equation is explicitly satisfied for this theory: 
\ie
\widehat{\bf Z}(\tau, \bar\tau) &= \begin{pmatrix} |\tau|^{{1\over2}} \sum_{n,m\in \mathbb Z}q^{\frac14\left(\frac nr + mr\right)^2} \bar{q}^{\frac14\left(\frac nr - mr\right)^2} \\ 0 \\ |\tau|^{{1\over2}} \sum_{n,m\in \mathbb Z}q^{\frac14\left(\frac nr + mr\right)^2} \bar{q}^{\frac14\left(\frac nr - mr\right)^2}  \end{pmatrix}
\fe
and therefore
\ie
 {\bf F} \, \widehat{\bf Z}(\tau, \bar\tau) =  \widehat{\bf Z}(\tau, \bar\tau) = \widehat{\bf Z}(-\frac1\tau, -\frac1{\bar\tau})
\fe

\section{Explicit form of the $\mathcal{N}=1$ Maloney-Witten partition function}
\label{sec:appmwk}

We aim for this Appendix to be self-contained.
We review the calculation of \cite{Maloney:2007ud}, first the bosonic case in Appendix~\ref{sec:MWK}, and then the $\mathcal{N}=1$ NS case in Appendix~\ref{sec:N1MWK}. We extend the calculation to the $\mathcal{N}=1$ Ramond sector in Appendix~\ref{sec:N1MWK} and look at sum of Ramond characters in both the NS and Ramond sectors in \ref{sec:N1MWKR}. We examine previously neglected consistencies of the results.

\subsection{Modular sum of bosonic character}
\label{sec:MWK}

Let us review the calculation done for the bosonic partition function \cite{Maloney:2007ud, Keller:2014xba}. The partition function is given by an expression
\be
Z^{\text{MWK}}(\tau,\bar\tau) = \sum_{\gamma \in\Gamma_\infty\backslash SL(2,\mathbb Z)} \chi_0(\gamma\tau) \chi_0(\gamma\bar\tau)
\label{eq:mwksum}
\ee
where $\chi_0(\tau)$ is the Virasoro vacuum character:
\be
\chi_0(\tau) = \frac{q^{-\frac{c-1}{24}}}{\eta(q)}(1-q).
\ee
The sum in (\ref{eq:mwksum}) diverges, but can be regulated to give a finite modular-invariant answer, as described in \cite{Maloney:2007ud}. The resulting partition has three unphysical features: 
\begin{enumerate}
\item The spectrum of primary operators is a continuous density of states rather than a discrete sum of delta functions for $h, \, \bar h > \frac{c-1}{24}$.
\item The degeneracy of primary operators is negative at $h=\bar h = \frac{c-1}{24}$.
\item The density of primary operators is negative at any odd spin, for twist $2\,\text{min}(h,\bar h)$ sufficiently close to $\frac{c-1}{12}$ \cite{Benjamin:2019stq}.
\end{enumerate}

To compute the actual sum in (\ref{eq:mwksum}), it is convenient to separate it into four terms coming from the product $(1-q)(1-\bar q) = 1 - q -\bar q + q\bar q$ necessary to subtract off null states. The sum (\ref{eq:mwksum}) over $SL(2,\mathbb Z)$ is done explicitly for each of these four terms in \cite{Maloney:2007ud, Keller:2014xba}. The first and last terms (where the ``seed" operator is a scalar), \cite{Keller:2014xba, Alday:2019vdr} pointed out that the result can be written conveniently as a sum of the ``non-holomorphic Eisenstein series"
\begin{align}
E(s, \tau, \bar\tau) &\equiv \sum_{\gamma\in \Gamma_\infty\backslash SL(2,\mathbb Z)} \text{Im}(\gamma\tau)^s \nn\\
&= y^s + \frac{\sqrt\pi\Gamma\(s-\frac12\) \zeta(2s-1)}{\Gamma(s)\zeta(2s)}y^{1-s} \nn
\\
& \hspace{.5in} + \sum_{j=1}^{\infty}\(e^{2\pi i j x} + e^{-2\pi i j x}\) \frac{2\pi^s\sigma_{2s-1}(j)}{\Gamma(s)\zeta(2s)j^{s-\frac12}} y^{\frac12} K_{s-\frac12}\(2\pi y j\),
\label{eq:EisenDef}
\end{align}
where
\be
\tau = x + i y, ~~~~~~\bar\tau=x-iy,
\ee
$\sigma$ is the divisor sigma function, and $K$ is the Bessel $K$ function. 
By construction, the non-holomorphic Eisenstein series satisfy
\be
E(s,\tau,\bar\tau) = E(s,\gamma\tau, \gamma\bar\tau), ~~~~~\gamma\in SL(2,\mathbb Z).
\ee
The regulated $SL(2,\mathbb Z)$ sum of a ``seed" term was computed in Section 3 of \cite{Keller:2014xba}.
We rewrite their results here as:
\be
\sum_{\gamma\in \Gamma_\infty \backslash SL(2,\mathbb Z)} \(\frac{q^{-\xi} \bar{q}^{-\xi}}{\left|\eta(\tau)\right|^2}\)_\gamma = \frac{1}{y^{{1\over2}} \left|\eta(\tau)\right|^2} \sum_{m=1}^{\infty} \frac{(4\pi \xi)^m E(m+\frac12, \tau,\bar\tau)}{m!}
\label{eq:Eisenseed}
\ee
and\footnote{The following identity for derivatives of Bessel $K$ functions
\begin{align}
y^{-m}\(\alpha_1 + \alpha_2 \frac{d}{dy}\)^m y^m K_m(\alpha_3 y) = \alpha_2^m \sum_{a=0}^m(-1)^{a+m}\frac{m!\alpha_3^{m-a}~_1F_1\(-a, 1-2a+m; -\frac{2\alpha_1 y}{\alpha_2}\)}{2^aa!(m-2a)!y^a} K_a\(\alpha_3 y\)
\label{eq:1f1bessel}
\end{align}
can be used to rewrite (\ref{eq:SpinningEisenseed}), and later (\ref{eq:scalarANDspinningNS}) and (\ref{eq:spinningRRR}) more explicitly.}
\begin{align}
&\sum_{\gamma\in \Gamma_\infty \backslash SL(2,\mathbb Z)} \(\frac{q^{-\xi+1} \bar{q}^{-\xi}}{\left|\eta(\tau)\right|^2}\)_\gamma = \frac{1}{\left|\eta(\tau)\right|^2} \Bigg[ e^{2\pi i x + 2\pi y (2\xi-1)} + 2 + \sum_{m=1}^{\infty} \frac{2\pi^{m+\frac12}T_m(2\xi-1)y^{-m}}{m\Gamma(m+\frac12)\zeta(2m+1)} \nn\\
&+ \sum_{\substack{j\neq0 \\ j \in \mathbb Z}} e^{2\pi i x j} \sum_{m=0}^{\infty} \(\sum_{s=1}^{\infty} \frac{S(j,1;s)}{s^{2m+1}}\) \frac{2^{3m+1} \pi^{2m}|j|^m}{(2m)! y^m} \((2\xi-1) + \frac{1}{2\pi j} \frac{d}{dy}\)^m y^m K_m(2\pi y |j|)
\Bigg].
\label{eq:SpinningEisenseed}
\end{align}
In (\ref{eq:SpinningEisenseed}), $T_m$ is the Chebyshev $T$ polynomial and $S(j,J;s)$ is a Kloosterman sum. The $m=0$ term in (\ref{eq:SpinningEisenseed}) needs to be defined via analytic continuation since sum over $s$ diverges.

\subsection{Modular sum of $\mathcal{N}=1$ NS character}
\label{sec:N1MWK}

Let us first define some notation when we discuss the $\mathcal{N}=1$ partition function. In this section we will focus on a modular sum of the NS vacuum character, and will look at the result obtained in both the NS and Ramond sectors. These will be denoted $Z^{\text{NS}-}\left[\ldots\right]$ and $Z^{\text{R}+}\left[\ldots\right]$ respectively, where the $\ldots$ represent what we are summing over. In this section we will compute $Z^{\text{NS}-}\left[|\chi_0^{\text{NS}}|^2\right]$ and $Z^{\text{R}+}\left[|\chi_0^{\text{NS}}|^2\right]$, and in Section \ref{sec:N1MWKR} we will generalize to looking at the modular sum of Ramond characters, and compute $Z^{\text{NS}-}\left[|\chi_\text{R}|^2\right]$ and $Z^{\text{R}+}\left[|\chi_\text{R}|^2\right]$.

The $\mathcal{N}=1$ NS partition function $Z^{\text{NS}-}\left[|\chi_0^{\text{NS}}|^2\right]$ is a $\Gamma^0(2)$ sum of the NS vacuum character. If we define the \emph{non-degenerate} character $\chi_0^{\text{non-deg}}$ as
\be
\chi_0^{\text{non-deg}}(\tau) = q^{-\frac{c-\frac32}{24}}\frac{\eta(\tau/2)}{\eta(\tau)^2}
\ee
then the vacuum state and descendants is given by
\be
|\chi_0^{\text{NS}-}(\tau)|^2 = |\chi_0^{\text{non-deg}}(\tau)|^2(1 + q^{{1\over2}} + \bar{q}^{{1\over2}} + q^{{1\over2}}\bar{q}^{{1\over2}})
\label{eq:716}
\ee 
(See (7.16) of \cite{Maloney:2007ud}.) Let us first focus on the first and last term of (\ref{eq:716}). Define the non-holomorphic Eisenstein series for the congruence subgroup $\Gamma^0(2)$ as
\begin{align}
E^{\Gamma^0(2)}\(s,\tau,\bar\tau\) &= y^s + \frac{2^{2s-1}\sqrt{\pi}\Gamma\(s-\frac12\) \zeta(2s-1)}{\(4^s-1\)\Gamma(s)\zeta(2s)}y^{1-s} \nn\\
&+ \sum_{j=1}^{\infty} \(e^{2\pi i j x} + e^{-2\pi i j x}\)\frac{2\pi^s\(4^s \sigma_{2s-1}\(j\) - \sigma_{2s-1}\(2j\)\)}{(4^s-1)\Gamma(s)\zeta(2s)j^{s-\frac12}} y^{\frac12} K_{s-\frac12}\(2\pi y j\) \nn\\
&- \sum_{j=\frac12, \frac32, \ldots} \(e^{2\pi i j x} + e^{-2\pi i j x}\)\frac{2\pi^s \sigma_{2s-1}\(2j\)}{(4^s-1)\Gamma(s)\zeta(2s) j^{s-\frac12}} y^\frac12 K_{s-\frac12}\(2\pi y j\).
\label{eq:eisenNS}
\end{align}
In terms of (\ref{eq:eisenNS}), the partition function $Z^{\text{NS}-}\left[|\chi_0^{\text{non-deg}}|^2\right]$ can be written compactly as
\be
\frac{y^{{1\over2}}|\eta(\tau)|^4}{|\eta(\tau/2)|^2}Z^{\text{NS}-}\left[|\chi_0^{\text{non-deg}}|^2\right](\tau,\bar\tau) = \sum_{m=1}^{\infty} \frac{(4\pi \xi)^m E^{\Gamma^0(2)}\(m+\frac12, \tau, \bar\tau\)}{m!},
\label{eq:MWKNS}
\ee
where $\xi = \frac{c-\frac32}{24}$. Similarly $Z^{\text{NS}-}\left[|\chi_0^{\text{non-deg}}|^2 q^{\frac12} \bar{q}^{\frac12}\right](\tau,\bar\tau)$ is given by the same expression but with $\xi$ replaced by $\xi+1 = \frac{c-\frac32}{24}+1$. 

The remaining terms in (\ref{eq:716}) are more complicated but can be calculated by modifying the techniques of \cite{Keller:2014xba}. The final answer for the remaining two terms is
\begin{align}
&\frac{|\eta(\tau)|^4}{|\eta(\tau/2)|^2}Z^{\text{NS}-}\left[|\chi_0^{\text{non-deg}}|^2 q^{\pm\frac12}\right](\tau,\bar\tau) = e^{\pm \pi i x + 2\pi y(2\xi-\frac12)}-2 - \sum_{m=1}^{\infty} \frac{2^{m+1} \pi^{m+\frac12} y^{-m} T_m(4\xi - 1)}{(2^{2m+1}-1)m\Gamma(m+\frac12)\zeta(2m+1)}  \nn\\
&+ \sum_{\substack{j\neq0 \\ 2j\in \mathbb Z}} e^{2\pi i j x} \sum_{m=0}^{\infty} \frac{2^{3m}\pi^{2m}}{(2m)!} \left| j \right|^m y^{-m} \(\sum_{s=1}^{\infty} \frac{S\(2j, \pm 1; 2s\)}{s^{2m+1}}\)\(2\xi-\frac12 \pm \frac{1}{4\pi j}\frac{d}{dy} \)^m y^m K_m(2\pi y |j|)
\label{eq:scalarANDspinningNS}
\end{align}
where the $m=0$ term needs to be defined via analytic continuation since the sum over $s$ diverges.
From (\ref{eq:MWKNS}) and (\ref{eq:scalarANDspinningNS}) we finally get the full $Z^{\text{NS}-}\left[|\chi_0^{\text{NS}}|^2\right]$.\footnote{To get $Z^{\text{NS}+}\left[|\chi_0^{\text{NS}}|^2\right]$, we simply change the signs of all half-integer spins.}

What is the corresponding Ramond sector partition function $Z^{\text{R}+}\left[|\chi_0^{\text{NS}}|^2\right]$? We can obtain this by doing an $S$ transformation on $Z^{\text{NS}-}\left[|\chi_0^{\text{NS}}|^2\right]$. The first step is to compute the $S$ transformation of $E^{\Gamma^0(2)}$. We get\footnote{This can be derived by doing the regularized sum over the coset $\Gamma^0(2)S$ (or equivalently $S\Gamma_0(2)$) instead of $\Gamma^0(2)$. We have also checked (\ref{eq:quickbrownfox}) numerically to extremely high precision.}
\begin{align}
E^{\Gamma^0(2)}\(s, -\frac1\tau,-\frac1{\bar\tau}\) &= \frac{(4^s-2)\sqrt \pi \Gamma\(s-\frac12\) \zeta(2s-1)}{(4^s-1)\Gamma(s)\zeta(2s)} y^{1-s} \nn\\ 
&+ \sum_{j=1}^{\infty} \(e^{2\pi i j x} + e^{-2\pi i j x}\)\frac{4\pi^s\(\sigma_{2s-1}(2j) - \sigma_{2s-1}(j)\) }{(4^s-1)\Gamma(s)\zeta(2s)j^{s-\frac12}} y^{\frac12} K_{s-\frac12}\(2\pi y j\).
\label{eq:quickbrownfox}
\end{align}
Under this definition we then have
\begin{equation}
\frac{y^{{1\over2}} |\eta(\tau)|^4}{2|\eta(2\tau)|^2}Z^{\text{R}+}\left[|\chi_0^{\text{non-deg}}|^2\right](\tau,\bar\tau) = \sum_{m=1}^{\infty}  \frac{(4\pi \xi)^m E^{\Gamma^0(2)}\(m+\frac12, -\frac1\tau, -\frac1{\bar\tau}\)}{m!}.
\label{eq:ZR}
\end{equation}
Note that the $y\rightarrow\infty$ limit of (\ref{eq:ZR}) is zero. Thus the first and fourth term in (\ref{eq:716}) give zero Ramond ground states. Finally note that
\be
\frac12\(E^{\Gamma^0(2)}(s, \tau,\bar\tau) + E^{\Gamma^0(2)}\(s, -\frac1\tau,-\frac1{\bar\tau}\) + E^{\Gamma^0(2)}(s, \tau+1,\bar\tau+1) \) = E(s, \tau,\bar\tau).
\label{eq:kindofgso}
\ee
This equality is because as we sum over the three terms in (\ref{eq:kindofgso}), we are effectively summing twice over all elements of $SL(2,\mathbb Z)$, which is the definition of $E(s, \tau, \bar\tau)$.  Twice because we sum over the elements of $SL(2,\mathbb Z)$ where we mod out by the group generated by $T^2$ on the left, not by the group generated by $T$. 

In principle we could compute the $S$ transform of (\ref{eq:scalarANDspinningNS}) to get $Z^{\text{R}+}\left[|\chi_0^{\text{non-deg}}|^2 q^{\pm\frac12}\right](\tau,\bar\tau)$. However, we can also use the same logic as in (\ref{eq:kindofgso}): The sum of the NS$+$, NS$-$, and R functions must give the $SL(2,\mathbb Z)$ sum over the term $|\chi_0^{\text{non-deg}}|^2 q^{\pm\frac12}$. Since this has non-integer spin, the $SL(2,\mathbb Z)$ sum vanishes, which allows us to obtain $Z^{\text{R}+}\left[|\chi_0^{\text{non-deg}}|^2 q^{\pm\frac12}\right](\tau,\bar\tau)$ from (\ref{eq:scalarANDspinningNS}). In particular\footnote{We have also numerically checked that the expression for $Z^{\text{R}+}\left[|\chi_0^{\text{non-deg}}|^2 q^{\pm\frac12}\right](\tau,\bar\tau)$ in (\ref{eq:spinningRRR}) is the $S$ transform of the expression for $Z^{\text{NS}-}\left[|\chi_0^{\text{non-deg}}|^2 q^{\pm\frac12}\right](\tau,\bar\tau)$ in (\ref{eq:scalarANDspinningNS}) and that both expressions are modular invariant under $\Gamma_0(2)$ and $\Gamma^0(2)$ respectively.}
\begin{align}
&\frac{|\eta(\tau)|^4}{2|\eta(2\tau)|^2}Z^{\text{R}+}\left[|\chi_0^{\text{non-deg}}|^2 q^{\pm\frac12}\right](\tau,\bar\tau) = 4 +  \sum_{m=1}^{\infty} \frac{2^{m+2} \pi^{m+\frac12} y^{-m} T_m(4\xi - 1)}{(2^{2m+1}-1)m\Gamma(m+\frac12)\zeta(2m+1)}  \nn\\
&- \sum_{\substack{j\neq0 \\ j\in \mathbb Z}} e^{2\pi i j x} \sum_{m=0}^{\infty} \frac{2^{3m+1}\pi^{2m}}{(2m)!} \left| j \right|^m y^{-m} \(\sum_{s=1}^{\infty} \frac{S\(2j, \pm 1; 2s\)}{s^{2m+1}}\)\(2\xi-\frac12 \pm \frac{1}{4\pi j}\frac{d}{dy} \)^m y^m K_m(2\pi y |j|).
\label{eq:spinningRRR}
\end{align}
Therefore each of the two terms in (\ref{eq:spinningRRR}) gives 8 Ramond sector ground states, which gives in total $16$ Ramond ground states.

Finally we would like to show that the Ramond sector is in fact positive everywhere. The relevant modular kernel $K^{\text{NS}\rightarrow\text{R}}(h)$ is
\begin{align}
\chi_0^{\text{NS}-}\(-1/\tau\) &= \int_{\frac c{24}}^\infty dh K^{\text{NS}\rightarrow\text{R}}(h) \chi^\text{R}(\tau) \nn\\
\end{align}
which is the first nontrivial element of the coset $\Gamma^0(2)S$. The kernels are given by
\be
K^{\text{NS}\rightarrow\text{R}}(h) =  \frac{1}{\sqrt{h-\frac c{24}}}\(\cosh\(4\pi\sqrt{\(\frac{c-\frac32}{24}\)\(h-\frac c{24}\)}\) + \cosh\(4\pi\sqrt{\(\frac{c-\frac{27}2}{24}\)\(h-\frac c{24}\)}\)\)
\label{eq:kkernels}
\ee
Crucially, (\ref{eq:kkernels}) scales as $\frac1{\sqrt{h-\frac c{24}}}$ in the limit as $h\rightarrow \frac{c}{24}$. This is to be contrasted with e.g. the Virasoro $S$ kernel which scales as $\sqrt{h-\frac{c-1}{24}}$ in the same limit. Because of the different scaling, we will not see the negativity in the high-spin, low-twist sector as in \cite{Benjamin:2019stq}, and the Ramond sector is positive everywhere. Finally we note that since this kernel has support at $h>\frac{c}{24}$ immediately, using the arguments of \cite{Collier:2016cls} this implies that any $\mathcal{N}=1$ SCFT with a nonzero twist gap in the NS sector has vanishing twist gap in the Ramond sector.

\subsection{Modular sum of $\mathcal{N}=1$ Ramond character}
\label{sec:N1MWKR}

We can try the same calculation but this time sum a state in the Ramond sector. The non-holomorphic Eisenstein series given by summing $y^s$ over elements of $\Gamma_0(2)$ is
\begin{align}
E^{\Gamma_0(2)}(s,\tau,\bar\tau) &= \sum_{\gamma \in \mathbb Z \backslash \Gamma_0(2)} y^s\Big|_\gamma
= y^s + \frac{\sqrt\pi \Gamma\(s-\frac12\) \zeta(2s-1)}{(4^s-1)\Gamma(s)\zeta(2s)}y^{1-s} \nn\\
& \hspace{-1in} + \sum_{j=1}^\infty\(e^{2\pi i j x} + e^{-2\pi i jx}\)\frac{(4^{s}+1)\sigma_{2s-1}(j) - 2\sigma_{2s-1}(2j)}{(4^s-1)j^{s-\frac12}\Gamma(s)\zeta(2s)}2\pi^s y^\frac12 K_{s-\frac12}\(2\pi y j\).
\label{eq:SGamma0}
\end{align}

The sum of a Ramond ground state with weight $h-\frac{c}{24} = \bar h - \frac{c}{24} = \kappa$ is
\be
\frac{y^{{1\over2}} |\eta(\tau)|^4}{2|\eta(2\tau)|^2}   Z^{\text{R}+}\left[|\chi_0^{\text{non-deg, R}}|^2 q^\kappa \bar{q}^\kappa \right](\tau,\bar\tau)  = \sum_{m=1}^{\infty} \frac{(-4\pi\kappa)^m E^{\Gamma_0(2)}(m+\frac12, \tau, \bar\tau)}{m!}.
\label{eq:REism}
\ee

The $S$-transform of (\ref{eq:SGamma0}) is given by
\begin{align}
E^{\Gamma_0(2)}\(s, -\frac1\tau,-\frac1{\bar\tau}\) &= \frac{(2^{2s-1}-1)\sqrt\pi \Gamma\(s-\frac12\)\zeta(2s-1)}{(4^s-1)\Gamma(s)\zeta(2s)}y^{1-s} \nn\\
& \hspace{-1in} + \sum_{j=1}^{\infty} \(e^{2\pi i j x} + e^{-2\pi i j x}\) \frac{2\pi^s \(\sigma_{2s-1}\(2j\) - \sigma_{2s-1}\(j\)\)}{(4^s-1)\Gamma(s)\zeta(2s) j^{s-\frac12}} y^\frac12 K_{s-\frac12}\(2\pi y j\) \nn\\
& \hspace{-1in} + \sum_{j=\frac12, \frac32, \ldots} \(e^{2\pi i j x} + e^{-2\pi i j x}\)\frac{2\pi^s \sigma_{2s-1}\(2j\)}{(4^s-1)\Gamma(s)\zeta(2s) j^{s-\frac12}} y^\frac12 K_{s-\frac12}\(2\pi y j\).
\end{align}
The sum of the three spin-structures again must give us $E(s, \tau, \bar\tau)$, and indeed
\be
E^{\Gamma_0(2)}\(s, \tau, \bar\tau\) + E^{\Gamma_0(2)}\(s, -\frac1\tau,-\frac1{\bar\tau}\) + E^{\Gamma_0(2)}\(s, -\frac1{\tau + 1},-\frac1{\bar\tau + 1}\) = E(s, \tau, \bar\tau). 
\label{eq:kingofgso}
\ee
Unlike in (\ref{eq:kindofgso}), the equality (\ref{eq:kingofgso}) has no relative factor of 2 since in $E^{\Gamma_0(2)}$ we mod out by the group generated by $T$ on the left (as opposed to $T^2)$. 

Thus NS sector coming from a modular sum of a Ramond sector scalar is given by
\be
\frac{y^{{1\over2}}|\eta(\tau)|^4}{|\eta(\tau/2)|^2}Z^{\text{NS}-}\left[|\chi_0^{\text{non-deg, R}}|^2 q^\kappa \bar{q}^\kappa \right](\tau,\bar\tau) = \sum_{m=1}^{\infty} \frac{(-4\pi \kappa)^m E^{\Gamma_0(2)}\(m+\frac12, -\frac1\tau,-\frac1{\bar\tau}\) }{m!}.
\label{eq:pigpigpigpig}
\ee

We end this Appendix by noting the following. In \cite{Alday:2019vdr}, a different regularization from \cite{Maloney:2007ud} was chosen to perform the $SL(2,\mathbb Z)$ sum of $\chi_0(\tau) \chi_0(\bar\tau)$ in the bosonic case. Although the intermediate steps of the calculation were complicated, the end result was that instead of the part of the partition function (\ref{eq:Eisenseed}) being given as a sum over non-holomorphic Eisenstein series of {\it half-integer} weight, it was given as a sum over non-holomorphic Eisenstein series of {\it integer} weight. It is natural to conjecture then that the analogous regularization scheme for pure $\mathcal{N}=1$ supergravity would simply give us (\ref{eq:MWKNS}) (\ref{eq:ZR}), (\ref{eq:REism}), (\ref{eq:pigpigpigpig}) with $m$ shifted by a ${1\over2}$.

\section{Reduced partition function and characters}
\label{App:Reduced}

In the practical implementation of modular bootstrap, it is convenient to work with reduced characters defined as\footnote{In defining the reduced characters, we have utilized some modular transformation properties of the Dedekind eta function.}
\ie
\widehat\chi_h^{\text{NS}}(\tau) &= \left( \frac{\eta(\tau)}{\eta(2\tau)\eta(\tau/2)} \right)^{-1} |\tau|^{1\over4} \times \chi_h^{\text{NS}}(\tau) \, ,
\\
\widehat\chi_h^{\text{NS},-}(\tau) &= \(\frac{\eta(\tau/2)}{\eta(\tau)^2}\)^{-1} |\tau|^{1\over4} \times \chi_h^{\text{NS},-}(\tau) \, , 
\\
\widehat\chi_h^{\text{R}}(\tau) &= {1 \over \sqrt2} \left( \frac{\eta(2\tau)}{\eta(\tau)^2} \right)^{-1} |\tau|^{1\over4} \times \chi_h^{\text{R}}(\tau) \, .
\fe
Explicitly, they are
\begin{align}
\widehat\chi_h^{\text{NS}}(\tau) &= \begin{cases}
\displaystyle q^{-\frac{c-\frac32}{24}}(1-q^{\frac12}) |\tau|^{{1\over4}} \qquad & h = 0
\\
\displaystyle q^{h-\frac{c-\frac32}{24}} |\tau|^{{1\over4}} \qquad & h > 0
\end{cases}
\\
\widehat\chi_h^{\text{NS},-}(\tau) &= \begin{cases}
\displaystyle q^{-\frac{c-\frac32}{24}}(1+q^{\frac12}) |\tau|^{{1\over4}} \qquad & h = 0
\\
\displaystyle e^{2\pi i h} q^{h-\frac{c-\frac32}{24}} |\tau|^{{1\over4}} \qquad & h > 0
\end{cases}
\end{align}
in the NS sector and
\begin{align}
\widehat\chi_h^{\text{R}}(\tau) = \begin{cases} \displaystyle \frac{1}{\sqrt2} |\tau|^{{1\over4}} & h = \frac{c}{24} 
\\ 
\displaystyle \sqrt{2} q^{h-\frac{c}{24}} |\tau|^{{1\over4}} \qquad & h > \frac c{24} \end{cases}
\end{align}
in the Ramond sector.  The reduced characters have the virtue that the non-degenerate NS+ and NS$-$ characters are now identical up to a phase.  The partition functions defined as sums over reduced characters transform properly under $S$.

\bibliographystyle{JHEP}
\bibliography{refs}

\providecommand{\href}[2]{#2}\begingroup\raggedright\begin{thebibliography}{10}

\bibitem{Belavin:1984vu}
A.~Belavin, A.~M. Polyakov, and A.~Zamolodchikov, {\it {Infinite Conformal
  Symmetry in Two-Dimensional Quantum Field Theory}},  {\em Nucl. Phys. B} {\bf
  241} (1984) 333--380.

\bibitem{Friedan:1983xq}
D.~Friedan, Z.-a. Qiu, and S.~H. Shenker, {\it {Conformal Invariance, Unitarity
  and Two-Dimensional Critical Exponents}},  {\em Phys. Rev. Lett.} {\bf 52}
  (1984) 1575--1578.

\bibitem{Cardy:1986ie}
J.~L. Cardy, {\it {Operator Content of Two-Dimensional Conformally Invariant
  Theories}},  {\em Nucl. Phys. B} {\bf 270} (1986) 186--204.

\bibitem{Hellerman:2009bu}
S.~Hellerman, {\it {A Universal Inequality for CFT and Quantum Gravity}},  {\em
  JHEP} {\bf 08} (2011) 130, [\href{http://arxiv.org/abs/0902.2790}{{\tt
  arXiv:0902.2790}}].

\bibitem{Keller:2012mr}
C.~A. Keller and H.~Ooguri, {\it {Modular Constraints on Calabi-Yau
  Compactifications}},  {\em Commun. Math. Phys.} {\bf 324} (2013) 107--127,
  [\href{http://arxiv.org/abs/1209.4649}{{\tt arXiv:1209.4649}}].

\bibitem{Friedan:2013cba}
D.~Friedan and C.~A. Keller, {\it {Constraints on 2d CFT partition functions}},
   {\em JHEP} {\bf 10} (2013) 180, [\href{http://arxiv.org/abs/1307.6562}{{\tt
  arXiv:1307.6562}}].

\bibitem{Qualls:2013eha}
J.~D. Qualls and A.~D. Shapere, {\it {Bounds on Operator Dimensions in 2D
  Conformal Field Theories}},  {\em JHEP} {\bf 05} (2014) 091,
  [\href{http://arxiv.org/abs/1312.0038}{{\tt arXiv:1312.0038}}].

\bibitem{Keller:2014xba}
C.~A. Keller and A.~Maloney, {\it {Poincare Series, 3D Gravity and CFT
  Spectroscopy}},  {\em JHEP} {\bf 02} (2015) 080,
  [\href{http://arxiv.org/abs/1407.6008}{{\tt arXiv:1407.6008}}].

\bibitem{Benjamin:2016fhe}
N.~Benjamin, E.~Dyer, A.~L. Fitzpatrick, and S.~Kachru, {\it {Universal Bounds
  on Charged States in 2d CFT and 3d Gravity}},  {\em JHEP} {\bf 08} (2016)
  041, [\href{http://arxiv.org/abs/1603.09745}{{\tt arXiv:1603.09745}}].

\bibitem{Kraus:2016nwo}
P.~Kraus and A.~Maloney, {\it {A cardy formula for three-point coefficients or
  how the black hole got its spots}},  {\em JHEP} {\bf 05} (2017) 160,
  [\href{http://arxiv.org/abs/1608.03284}{{\tt arXiv:1608.03284}}].

\bibitem{Collier:2016cls}
S.~Collier, Y.-H. Lin, and X.~Yin, {\it {Modular Bootstrap Revisited}},  {\em
  JHEP} {\bf 09} (2018) 061, [\href{http://arxiv.org/abs/1608.06241}{{\tt
  arXiv:1608.06241}}].

\bibitem{Collier:2017shs}
S.~Collier, P.~Kravchuk, Y.-H. Lin, and X.~Yin, {\it {Bootstrapping the
  Spectral Function: On the Uniqueness of Liouville and the Universality of
  BTZ}},  {\em JHEP} {\bf 09} (2018) 150,
  [\href{http://arxiv.org/abs/1702.00423}{{\tt arXiv:1702.00423}}].

\bibitem{Bae:2017kcl}
J.-B. Bae, S.~Lee, and J.~Song, {\it {Modular Constraints on Conformal Field
  Theories with Currents}},  {\em JHEP} {\bf 12} (2017) 045,
  [\href{http://arxiv.org/abs/1708.08815}{{\tt arXiv:1708.08815}}].

\bibitem{Dyer:2017rul}
E.~Dyer, A.~L. Fitzpatrick, and Y.~Xin, {\it {Constraints on Flavored 2d CFT
  Partition Functions}},  {\em JHEP} {\bf 02} (2018) 148,
  [\href{http://arxiv.org/abs/1709.01533}{{\tt arXiv:1709.01533}}].

\bibitem{Anous:2018hjh}
T.~Anous, R.~Mahajan, and E.~Shaghoulian, {\it {Parity and the modular
  bootstrap}},  {\em SciPost Phys.} {\bf 5} (2018), no.~3 022,
  [\href{http://arxiv.org/abs/1803.04938}{{\tt arXiv:1803.04938}}].

\bibitem{Bae:2018qym}
J.-B. Bae, S.~Lee, and J.~Song, {\it {Modular Constraints on Superconformal
  Field Theories}},  {\em JHEP} {\bf 01} (2019) 209,
  [\href{http://arxiv.org/abs/1811.00976}{{\tt arXiv:1811.00976}}].

\bibitem{Afkhami-Jeddi:2019zci}
N.~Afkhami-Jeddi, T.~Hartman, and A.~Tajdini, {\it {Fast Conformal Bootstrap
  and Constraints on 3d Gravity}},  {\em JHEP} {\bf 05} (2019) 087,
  [\href{http://arxiv.org/abs/1903.06272}{{\tt arXiv:1903.06272}}].

\bibitem{Lin:2019kpn}
Y.-H. Lin and S.-H. Shao, {\it {Anomalies and Bounds on Charged Operators}},
  {\em Phys. Rev.} {\bf D100} (2019), no.~2 025013,
  [\href{http://arxiv.org/abs/1904.04833}{{\tt arXiv:1904.04833}}].

\bibitem{Mukhametzhanov:2019pzy}
B.~Mukhametzhanov and A.~Zhiboedov, {\it {Modular invariance, tauberian
  theorems and microcanonical entropy}},  {\em JHEP} {\bf 10} (2019) 261,
  [\href{http://arxiv.org/abs/1904.06359}{{\tt arXiv:1904.06359}}].

\bibitem{Hartman:2019pcd}
T.~Hartman, D.~Mazac, and L.~Rastelli, {\it {Sphere Packing and Quantum
  Gravity}},  {\em JHEP} {\bf 12} (2019) 048,
  [\href{http://arxiv.org/abs/1905.01319}{{\tt arXiv:1905.01319}}].

\bibitem{Pal:2019zzr}
S.~Pal and Z.~Sun, {\it {Tauberian-Cardy formula with spin}},  {\em JHEP} {\bf
  01} (2020) 135, [\href{http://arxiv.org/abs/1910.07727}{{\tt
  arXiv:1910.07727}}].

\bibitem{Pal:2020aa}
S.~Pal and Z.~Sun, {\it {High Energy Modular Bootstrap, Global Symmetries and
  Defects}},  \href{http://arxiv.org/abs/2004.12557}{{\tt arXiv:2004.12557}}.

\bibitem{Gliozzi:1976qd}
F.~Gliozzi, J.~Scherk, and D.~I. Olive, {\it {Supersymmetry, Supergravity
  Theories and the Dual Spinor Model}},  {\em Nucl. Phys.} {\bf B122} (1977)
  253--290.

\bibitem{Gaiotto:2015zta}
D.~Gaiotto and A.~Kapustin, {\it {Spin TQFTs and fermionic phases of matter}},
  {\em Int. J. Mod. Phys. A} {\bf 31} (2016), no.~28n29 1645044,
  [\href{http://arxiv.org/abs/1505.05856}{{\tt arXiv:1505.05856}}].

\bibitem{Bhardwaj:2016clt}
L.~Bhardwaj, D.~Gaiotto, and A.~Kapustin, {\it {State sum constructions of
  spin-TFTs and string net constructions of fermionic phases of matter}},  {\em
  JHEP} {\bf 04} (2017) 096, [\href{http://arxiv.org/abs/1605.01640}{{\tt
  arXiv:1605.01640}}].

\bibitem{Kapustin:2017jrc}
A.~Kapustin and R.~Thorngren, {\it {Fermionic SPT phases in higher dimensions
  and bosonization}},  {\em JHEP} {\bf 10} (2017) 080,
  [\href{http://arxiv.org/abs/1701.08264}{{\tt arXiv:1701.08264}}].

\bibitem{Thorngren:2018bhj}
R.~Thorngren, {\it {Anomalies and Bosonization}},
  \href{http://arxiv.org/abs/1810.04414}{{\tt arXiv:1810.04414}}.

\bibitem{Gaiotto:2018ypj}
D.~Gaiotto and T.~Johnson-Freyd, {\it {Holomorphic SCFTs with small index}},
  \href{http://arxiv.org/abs/1811.00589}{{\tt arXiv:1811.00589}}.

\bibitem{Karch:2019lnn}
A.~Karch, D.~Tong, and C.~Turner, {\it {A Web of 2d Dualities: ${\bf Z}_2$
  Gauge Fields and Arf Invariants}},  {\em SciPost Phys.} {\bf 7} (2019) 007,
  [\href{http://arxiv.org/abs/1902.05550}{{\tt arXiv:1902.05550}}].

\bibitem{yujitasi}
Y.~Tachikawa, ``{TASI 2019 Lectures}.''
  \url{https://member.ipmu.jp/yuji.tachikawa/lectures/2019-top-anom}.

\bibitem{Kaidi:2020aa}
J.~Kaidi, J.~Parra-Martinez, and Y.~Tachikawa, {\it {GSO projections via SPT
  phases}},  {\em Phys. Rev. Lett.} {\bf 124} (2020), no.~12 121601,
  [\href{http://arxiv.org/abs/1908.04805}{{\tt arXiv:1908.04805}}].

\bibitem{Ji:2019ugf}
W.~Ji, S.-H. Shao, and X.-G. Wen, {\it {Topological Transition on the Conformal
  Manifold}},  \href{http://arxiv.org/abs/1909.01425}{{\tt arXiv:1909.01425}}.

\bibitem{Hsin:2019gvb}
P.-S. Hsin and S.-H. Shao, {\it {Lorentz Symmetry Fractionalization and
  Dualities in (2+1)d}},  \href{http://arxiv.org/abs/1909.07383}{{\tt
  arXiv:1909.07383}}.

\bibitem{Hsieh:2020aa}
C.-T. Hsieh, Y.~Nakayama, and Y.~Tachikawa, {\it {On fermionic minimal
  models}},  \href{http://arxiv.org/abs/2002.12283}{{\tt arXiv:2002.12283}}.

\bibitem{Kulp:2020aa}
J.~Kulp, {\it {Two More Fermionic Minimal Models}},
  \href{http://arxiv.org/abs/2003.04278}{{\tt arXiv:2003.04278}}.

\bibitem{Witten:2007kt}
E.~Witten, {\it {Three-Dimensional Gravity Revisited}},
  \href{http://arxiv.org/abs/0706.3359}{{\tt arXiv:0706.3359}}.

\bibitem{Benjamin_2016}
N.~Benjamin, E.~Dyer, A.~L. Fitzpatrick, A.~Maloney, and E.~Perlmutter, {\it
  {Small Black Holes and Near-Extremal CFTs}},  {\em JHEP} {\bf 08} (2016) 023,
  [\href{http://arxiv.org/abs/1603.08524}{{\tt arXiv:1603.08524}}].

\bibitem{Maloney:2007ud}
A.~Maloney and E.~Witten, {\it {Quantum Gravity Partition Functions in Three
  Dimensions}},  {\em JHEP} {\bf 02} (2010) 029,
  [\href{http://arxiv.org/abs/0712.0155}{{\tt arXiv:0712.0155}}].

\bibitem{Hoehn}
G.~Hoehn, {\it Selbstduale vertexoperatorsuperalgebren und das babymonster
  (self-dual vertex operator super algebras and the baby monster)},  {\em
  Bonner Mathematische Schriften} {\bf 286} (1996) 1--85,
  [\href{http://arxiv.org/abs/0706.0236}{{\tt arXiv:0706.0236}}].

\bibitem{Benjamin:2019stq}
N.~Benjamin, H.~Ooguri, S.-H. Shao, and Y.~Wang, {\it {Light-cone modular
  bootstrap and pure gravity}},  {\em Phys.\ Rev.\ D} {\bf 100} (2019), no.~6
  066029, [\href{http://arxiv.org/abs/1906.04184}{{\tt arXiv:1906.04184}}].

\bibitem{Saad:2019lba}
P.~Saad, S.~H. Shenker, and D.~Stanford, {\it {JT gravity as a matrix
  integral}},  \href{http://arxiv.org/abs/1903.11115}{{\tt arXiv:1903.11115}}.

\bibitem{Gel_fand_1985}
I.~M. Gel'fand and A.~V. Zelevinskii, {\it Models of representations of
  classical groups and their hidden symmetries},  {\em Functional Analysis and
  Its Applications} {\bf 18} (1985), no.~3 183--198.

\bibitem{Shtepin_1994}
V.~V. Shtepin, {\it Intermediate lie algebras and their finite-dimensional
  representations},  {\em Russian Academy of Sciences. Izvestiya Mathematics}
  {\bf 43} (Jun, 1994) 559--579.

\bibitem{Landsberg:aa}
J.~M. Landsberg and L.~Manivel, {\it The sextonions and $e_{7\frac 12}$},  {\em
  Advances in Mathematics} {\bf 201} (2006) 143--179,
  [\href{http://arxiv.org/abs/math/0402157}{{\tt math/0402157}}].

\bibitem{Kawasetsu:2013}
K.~Kawasetsu, {\it {The Intermediate Vertex Subalgebras of the Lattice Vertex
  Operator Algebras}},  {\em Lett. Math. Phys.} {\bf 104} (2014), no.~2
  157--178, [\href{http://arxiv.org/abs/1305.6463}{{\tt arXiv:1305.6463}}].

\bibitem{Nazarov:2011mv}
A.~Nazarov, {\it {Affine.m - Mathematica package for computations in
  representation theory of finite-dimensional and affine Lie algebras}},  {\em
  Comput. Phys. Commun.} {\bf 183} (2012) 2480--2493,
  [\href{http://arxiv.org/abs/1107.4681}{{\tt arXiv:1107.4681}}].

\bibitem{Rattazzi:2008pe}
R.~Rattazzi, V.~S. Rychkov, E.~Tonni, and A.~Vichi, {\it {Bounding scalar
  operator dimensions in 4D CFT}},  {\em JHEP} {\bf 12} (2008) 031,
  [\href{http://arxiv.org/abs/0807.0004}{{\tt arXiv:0807.0004}}].

\bibitem{Simmons-Duffin:2015qma}
D.~Simmons-Duffin, {\it {A Semidefinite Program Solver for the Conformal
  Bootstrap}},  {\em JHEP} {\bf 06} (2015) 174,
  [\href{http://arxiv.org/abs/1502.02033}{{\tt arXiv:1502.02033}}].

\bibitem{Landry:2019qug}
W.~Landry and D.~Simmons-Duffin, {\it {Scaling the semidefinite program solver
  SDPB}},  \href{http://arxiv.org/abs/1909.09745}{{\tt arXiv:1909.09745}}.

\bibitem{Lin2019BoundsOT}
Y.-H. Lin, D.~Meltzer, S.-H. Shao, and A.~Stergiou, {\it {Bounds on Triangle
  Anomalies in (3+1)$d$}},  \href{http://arxiv.org/abs/1909.11676}{{\tt
  arXiv:1909.11676}}.

\bibitem{Alavirad2019AnomaliesAU}
Y.~Alavirad and M.~Barkeshli, {\it {Anomalies and unnatural stability of
  multi-component Luttinger liquids in $\mathbb{Z}_n\times\mathbb{Z}_n$ spin
  chains}},  \href{http://arxiv.org/abs/1910.00589}{{\tt arXiv:1910.00589}}.

\bibitem{Beem_2015}
C.~Beem, M.~Lemos, P.~Liendo, W.~Peelaers, L.~Rastelli, and B.~C. van Rees,
  {\it {Infinite Chiral Symmetry in Four Dimensions}},  {\em Commun. Math.
  Phys.} {\bf 336} (2015), no.~3 1359--1433,
  [\href{http://arxiv.org/abs/1312.5344}{{\tt arXiv:1312.5344}}].

\bibitem{Lemos:2015orc}
M.~Lemos and P.~Liendo, {\it {$\mathcal{N}=2$ central charge bounds from $2d$
  chiral algebras}},  {\em JHEP} {\bf 04} (2016) 004,
  [\href{http://arxiv.org/abs/1511.07449}{{\tt arXiv:1511.07449}}].

\bibitem{Beem:2018duj}
C.~Beem, {\it {Flavor Symmetries and Unitarity Bounds in ${\mathcal N}=2$
  Superconformal Field Theories}},  {\em Phys. Rev. Lett.} {\bf 122} (2019),
  no.~24 241603, [\href{http://arxiv.org/abs/1812.06099}{{\tt
  arXiv:1812.06099}}].

\bibitem{Alday:2019vdr}
L.~F. Alday and J.-B. Bae, {\it {Rademacher Expansions and the Spectrum of 2d
  CFT}},  \href{http://arxiv.org/abs/2001.00022}{{\tt arXiv:2001.00022}}.

\end{thebibliography}\endgroup

\end{document}